\documentclass[12pt,english]{article}
\usepackage[T1]{fontenc}
\usepackage[latin1]{inputenc}
\usepackage{a4}
\usepackage{babel}
\usepackage{graphics}
\usepackage{setspace}
\usepackage{subfigure}

\makeatletter

\providecommand{\LyX}{L\kern-.1667em\lower.25em\hbox{Y}\kern-.125emX\@}

\makeatother
\begin{document}

\title{\textbf{Stochastic Simulation of Gene Expression in a Single Cell }}

\author{Indrani Bose, Rajesh Karmakar and Siddhartha Roy\( ^{*} \) }

\maketitle
{\centering Department of Physics and \( ^{*} \) Department of Biophysics\par}

{\centering Bose Institute\par}

{\centering 93/1, A.P.C. Road, Kolkata-700 009,\par}

{\centering \( ^{*} \) P1/12, C.I.T. Scheme VII M, Kolkata-700 054,\par}

{\centering India.\par}

\begin{abstract}
In this paper, we consider two stochastic models of gene expression
in prokaryotic cells. In the first model, sixteen biochemical reactions
involved in transcription, translation and transcriptional regulation
in the presence of inducer molecules are considered. The time evolution
of the number of biomolecules of a particular type is determined using
the stochastic simulation method based on the Gillespie Algorithm.
The results obtained show that if the number of inducer molecules,
\( N_{I} \), is greater than or equal to the number of regulatory
molecules, \( N_{R}, \) the average protein level is high in the
steady state (state 2). The magnitude of the level is the same as
long as \( N_{I} \) \( \geq  \) \( N_{R}. \) When \( N_{I} \)
is \( \ll  \) \( N_{R} \), the average protein level is low, practically
zero (state 1). As \( N_{I} \) increases, the protein level continues
to remain low. When \( N_{I} \) becomes close to \( N_{R} \), protein
levels in the steady state are intermediate between high and low.
In the presence of autocatalysis, a cell mostly exists in either state
1 or state 2 giving rise to a bimodal distribution in the protein
levels in an ensemble of cells. This corresponds to the {}``all or
none'' phenomenon observed in experiments. In the second model, the
inducer molecules are not considered explicitly. An exhaustive simulation
over the parameter space of the model shows that there are three major
patterns of gene expression, Type A, Type B and Type C. The effect
of varying the cellular parameters on the patterns, in particular,
the transition from one type of pattern to another, is studied. Type
A and Type B patterns have been observed in experiments. Simple mathematical
models of transcriptional regulation predict Type C pattern of gene
expression in certain parameter regimes. The model studied by us includes
all the major biochemical reactions involved in gene expression and
the stochastic simulation results provide an understanding of the
microscopic origins of the different patterns of gene expression.
\end{abstract}

\section*{I. Introduction}

Gene expression is the central activity in a living cell. Genes are
fragments of DNA molecules and provide the blueprint for the synthesis
of functional molecules such as RNAs and proteins. In each cell, at
any instant of time, only a subset of genes present is active in directing
RNA/protein synthesis. The gene expression is {}``on'' in such a
case. There are two major steps in gene expression: transcription
and translation. During transcription the sequence along one of the
strands of the DNA molecule is copied or transcribed in a RNA molecule
(mRNA). During translation, the sequence of the mRNA molecule is translated
into the sequence of amino acids constituting a protein, i.e., a protein
molecule is synthesized. Regulation of gene expression is an essential
process in the living cell and determines the rates and patterns of
gene expression. An in-depth understanding of gene expression and
its regulation is the central focus of biology \cite{key-1}. 

The processes of transcription and translation involve several biochemical
reactions, the kinetics of which determine how the number of participating
biomolecules changes as a function of time. In the traditional differential
rate-equation approach, the time evolution of a system of chemical
reactions is assumed to be continuous and deterministic. In reality,
the time evolution is not a continuous process as molecular population
levels in a reacting system change only by discrete integer amounts.
Furthermore, the time evolution is not deterministic as the collision
of molecules which brings about a chemical reaction is a probabilistic
event. The deterministic rate equation approach is justified when
the number of molecules of each chemical species is large compared
to thermal fluctuations in the concentration. In a living cell, the
number of molecules participating in different biochemical reactions
is often small and there are considerable fluctuations in the reaction
rates. As a result, the time evolution of the reacting system, in
terms of how the number of reacting molecules changes as a function
of time is stochastic rather than deterministic. There is now an increasing
realization that stochasticity plays an important role in determining
the outcome of biochemical processes in the cell \cite{key-2}. Stochastic
effects in gene expression explain the pronounced cell-cell variation
observed in isogenic populations. A cell may have the option of proceeding
along one of two possible developmental pathways. The pathway selection
is probabilistic and the cell fate depends on the particular choice
of pathway. Thus, even a clonal population of cells can give rise
to two distinct subpopulations in the course of time. The randomization
of pathway choice leads to diversity and increases the likelihood
of survival of organisms in widely different environments.

The issue of stochasticity (randomness or noise) and its effect on
cellular processes as well as on the operation of synthetic devices
like genetic switches, has been addressed in several theoretical studies
\cite{key-4,key-6,key-7,key-8,key-9,key-10,key-11,key-12,key-13}.
A complete understanding of cellular processes requires an appreciation
of events at the level of an individual cell and subsequent extrapolation
to an ensemble of cells. Recent experimental advances have made it
possible to study processes within a single cell unmasked by ensemble
averaging \cite{key-15}. The simplest event one can study at the
single cell level is that of the expression of a reporter gene such
as \emph{lacZ} and \emph{GFP}. In the former case, the end product
is an enzyme \( \beta  \)-galactosidase which is capable of hydrolyzing
a noncolored substrate to a colored product. In the latter case, the
protein itself is fluorescent. Hence, the gene expression can be directly
studied either colorimetrically or fluorometrically at the level of
an individual cell. Recent experiments using such techniques, provide
evidence that gene expression occurs in abrupt stochastic bursts at
the level of an individual cell \cite{key-16,key-18,key-19}. Two
very recent experiments \cite{key-20,key-21} provide direct evidence
of stochasticity in gene expression. In both the experiments, a quantitative
measure of the noise associated with gene expression has been obtained.
The noise has both intrinsic and extrinsic components. Intrinsic noise
is the difference in protein synthesis which arises when two identical
copies of a gene are expressed under the same conditions. Extrinsic
noise occurs due to fluctuations in the cellular components required
for gene expression. The experiments provide a quantitative framework
for the characterization of noise in gene regulatory networks. 

Some earlier experiments on both prokaryotic and eukaryotic cells
have provided evidence of the so-called {}``all or none'' phenomenon
in gene expression \cite{key-16,key-18,key-19,key-22,key-23,key-24,key-56}. This
implies that in an individual cell, gene expression is either low/off
or has a high value. In an ensemble of cells the protein levels are
distributed in a bimodal manner, a large fraction of cells synthesize
proteins at a low (may be zero) level or produce them at a high level.
Most of the experiments require the presence of inducers/enhancers
to observe bimodality. There is strong experimental evidence that
inducer/enhancers increase the number of expressing cells but not
the level of expression per cell \cite{key-25}. The process of gene
expression is analogous to a binary switch which can be in {}``on''
and {}``off'' positions. Inducer/enhancer molecules make it favorable
for the switch to be in the {}``on'' position. Some theories have
been proposed so far to explain the so-called {}``all or none''
phenomenon in prokaryotic gene expression. The theories are mostly
based on an autocatalytic feedback mechanism \cite{key-22,key-23,key-24,key-27},
synthesis of the gene product gives rise to the transport or production
of inducer molecules which in turn promote further gene expression.
In section II of this paper, we propose a model of gene expression
and show that in the presence of a sufficient number of inducer molecules
in a cell, gene expression in that cell occurs at a high level. In
cells where inducer molecules are absent or are few in number, gene
expression occurs at practically zero level. The role of autocatalysis
in the {}``all or none'' phenomenon is also commented upon. The
method employed for the study is that of stochastic simulation based
on the Gillespie Algorithm (GA) \cite{key-28}. The GA provides a
stochastic realization of the temporal pattern of gene expression
and is more realistic and accurate than the deterministic differential
rate equation approach. Our model of gene expression includes the
major biochemical reactions involved in transcription and translation.
In section III, we explore the parameter space of the model and obtain
different temporal patterns of gene expression. One parameter region
of particular interest corresponds to stochastic flips between the
states 1 and 2 at random time intervals. In state 1, the protein level
is zero. This is an example of a binary switch which makes stochastic
transitions between the states 1 and 2 and the temporal process is
analogous to a two-state jump phenomenon. The effect of changing the
reaction parameters on the temporal patterns of gene expression is
further studied. Some of the results can be understood in the framework
of a simple mathematical model.

\section*{II. Stochastic model of gene expression}

We consider a single gene. The gene is transcribed into mRNA by an
enzyme called \( \textrm{RNA} \) polymerase (\( \textrm{RNAP} \)).
The process is initiated with the binding of RNAP to a site called
promoter, usually near the beginning of the transcribed sequence.
Expression of most genes are regulated at the level of transcription
and more specifically during the initiation of transcription, that
is, before the first phosphodiester bond is formed. Regulation of
transcription initiation is achieved by the binding of a regulatory
protein (\( \textrm{R} \)) to an overlapping segment of DNA (called
operator \( \textrm{O} \)) resulting in a turning off of mRNA production.
RNAP and the regulatory R molecules are mutually exclusive. If RNAP
binds to the promoter region first, it prevents the binding of R to
the operator region and vice versa. An inducer molecule (\( \textrm{I} \))
may bind to R both when R is free and when R is bound to the operator
O. In the later case, the complex of I and R detaches from the operator.
As long as R is forming a bound complex with I, it is unable to bind
at O and so cannot function as a regulatory protein. R regains its
activity when the inducer dissociates from the I\_R complex and R
is able to occupy the operator region once more. The biochemical reactions
considered in the model of gene expression are:\\
Reaction 1: \begin{equation}
\label{mathed:first-eqn}
\textrm{O}+\textrm{R}=\textrm{O}_{-}\textrm{R}
\end{equation}
Regulatory molecule R binds to the operator region O to form the bound
complex \( \textrm{O}_{-}\textrm{R} \).\\
Reaction 2:\begin{equation}
\label{mathed:second-eqn}
\textrm{O}_{-}\textrm{R}\rightarrow \textrm{O}+\textrm{R}
\end{equation}
Bound complex O\_R dissociates into free R and O.\\
Reaction 3:\begin{equation}
\label{mathed:third-eqn}
\textrm{P}+\textrm{RNAP}=\textrm{P}_{-}\textrm{RNAP}_{cc}
\end{equation}
RNA polymerase (RNAP) binds to the promoter region P forming the closed
complex P\_RNAP\( _{cc} \).\\
Reaction 4:\begin{equation}
\label{mathed:fourth-eqn}
\textrm{P}_{-}\textrm{RNAP}_{cc}\rightarrow \textrm{P}+\textrm{RNAP}
\end{equation}
The closed complex dissociates into free RNAP and P.\\
Reaction 5: \begin{equation}
\label{mathed:fifth-eqn}
\textrm{P}_{-}\textrm{RNAP}_{cc}\rightarrow \textrm{P}_{-}\textrm{RNAP}_{oc}
\end{equation}
Isomerization of closed to open complex P\_RNAP\( _{oc} \) occurs.
The open complex is the activated form of the RNAP-promoter complex.\\
Reaction 6:\begin{equation}
\label{mathed:sixth-eqn}
\textrm{P}_{-}\textrm{RNAP}_{oc}\rightarrow \textrm{TrRNAP}+\textrm{RBS}+\textrm{P}
\end{equation}
RNAP clears the promoter region and synthesis of the mRNA chain starts.
The appearance of the ribosome binding site RBS occurs at the beginning
of the mRNA chain. TrRNAP denotes transcribing RNA polymerase.\\
Reaction 7: \begin{equation}
\label{mathed:seventh-eqn}
\textrm{TrRNAP}\rightarrow \textrm{RNAP}
\end{equation}
RNAP completes transcription and is released from DNA.\\
Reaction 8:\begin{equation}
\label{mathed:eigth-eqn}
\textrm{RBS}+\textrm{Ribosome}\rightarrow \textrm{RibRBS}
\end{equation}
Ribosome binds to RBS and RibRBS denotes the bound complex.\\
Reaction 9:\begin{equation}
\label{mathed:nineth-eqn}
\textrm{RibRBS}\rightarrow \textrm{RBS}+\textrm{Ribosome}
\end{equation}
Ribosome dissociates from the bound complex RibRBS. \\
Reaction 10:\begin{equation}
\label{mathed:tenth-eqn}
\textrm{RBS}\rightarrow \textrm{degradation}
\end{equation}
RBS degrades due to the binding of RNAseE at RBS. This binding event
is not considered separately.\\
Reaction 11:\begin{equation}
\label{mathed:eleventh-eqn}
\textrm{RibRBS}\rightarrow \textrm{EIRib}+\textrm{RBS}
\end{equation}
RBS is cleared and the ribosome EIRib initiates translation of mRNA
chain.\\
Reaction 12:\begin{equation}
\label{mathed:twelveth-eqn}
\textrm{EIRib}\rightarrow \textrm{protein}
\end{equation}
Protein synthesis by transcribing ribosome is completed.\\
Reaction 13:\begin{equation}
\label{mathed:thirteenth-eqn}
\textrm{protein}\rightarrow \textrm{degradation}
\end{equation}
Degradation of protein product occurs.\\
Reaction 14:\begin{equation}
\label{mathed:fourteenth-eqn}
\textrm{I}+\textrm{R}\rightarrow \textrm{I}_{-}\textrm{R}
\end{equation}
Inducer molecule binds to free regulatory molecule R. I\_R is the
bound complex of I and R.\\
Reaction 15:\begin{equation}
\label{mathed:fifteenth-eqn}
\textrm{I}_{-}\textrm{R}\rightarrow \textrm{I}+\textrm{R}
\end{equation}
Bound complex dissociates into free I and R.\\
Reaction 16:\begin{equation}
\label{mathed:sixteenth-eqn}
\textrm{O}_{-}\textrm{R}+\textrm{I}\rightarrow \textrm{I}_{-}\textrm{R}+\textrm{O}
\end{equation}
Inducer binds to bound complex \( \textrm{O}_{-}\textrm{R} \), the
complex \( \textrm{I}_{-}\textrm{R} \) detaches and the operator
region O is freed. We emphasize that for many of the steps described
above, alternate mechanisms exist. However, the mechanism described
here is consistent with many gene regulatory systems.

Reaction schemes (1) - (13) are based on those considered in Refs.\cite{key-6,key-11}.
Transcription and translation are tightly coupled in prokaryotes.
As soon as RNAP leaves the promoter region, the 5\( ^{\prime } \)
end of the mRNA chain, containing the RBS, is available for ribosome
binding (Reaction 6). This implies that the mRNA chain need not be
completely synthesized to allow for ribosome binding (Reaction 8),
protein synthesis by translating ribosome (Reaction 12) and mRNA degradation
at RBS (Reaction 10). Following Ref.\cite{key-11}, the number of
mRNA molecules at any instant of time is given by the sum of the numbers
of RBS and RibRBS, the bound complex of ribosome and RBS. The functional
degradation of mRNA starts at the moment RNAseE binds to the RBS. 

We now give a brief description of the Gillespie Algorithm (GA) \cite{key-28}.
Suppose there are N chemical species participating in M chemical reactions.
Let X(i), i = 1, 2, 3 . . . . , N denote the number of molecules of
the i th chemical species. Given the values of X(i), i = 1, 2, 3,
. . . , N at a time \( t \), the GA is designed to answer two questions:
(1) when will the next reaction occur? and (2) what type of reaction
will it be? Let the next reaction occur at time \( t \)\( + \)\( \tau  \).
Knowing the type of reaction, one can adjust the numbers of participating
molecules in accordance with the reaction schemes. Thus, with repeated
applications of the GA, one can keep track of how the numbers X(i)'s
change as a function of time due to the occurrence of M different
types of chemical reactions. Each reaction \( \mu  \) (\( \mu  \)
= 1, 2, . . . , M) has a stochastic rate constant \( c \)\( _{\mu } \)
associated with it. This rate constant has the following interpretation: 

\( c \)\( _{\mu } \)\( dt \) = probability that a particular combination
of reactant molecules participates in the \( \mu  \) th reaction
in the infinitesimal time interval (\( t \), \( t+dt \)). Let \( h \)\( _{\mu } \)
be the number of distinct molecular combinations for the \( \mu  \)
th reaction. Then

\( a \)\( _{\mu } \)\( dt \)\( = \)\( h \)\( _{\mu } \)\( c \)\( _{\mu } \)\( dt \)
= probability that the \( \mu  \) th reaction occurs in the infinitesimal
time interval (\( t \), \( t+dt \)). \\
Let P(\( \tau  \),\( \mu  \)) \( d \)\( \tau  \) be the probability
that the next reaction is of type \( \mu  \) and occurs in the time
interval (\( t \)+ \( \tau  \), \( t \)+\( \tau  \) + d\( \tau  \)).
It is straightforward to show that\begin{equation}
\label{mathed:seventeenth-eqn}
\textrm{P}(\tau ,\mu )=a_{\mu }\exp (-a_{0}\tau )
\end{equation}
where \begin{equation}
\label{mathed:nineteenth-eqn}
a_{0}=\sum ^{\textrm{M}}_{\nu =1}a_{\nu }\equiv \sum ^{\textrm{M}}_{\nu =1}h_{\nu }c_{\nu }
\end{equation}

What is needed now is to generate a random pair (\( \tau  \),\( \mu  \))
according to the probability distribution (17). Let r\( _{1} \) and
r\( _{2} \) be two random numbers obtained by invoking the standard
unit interval uniform random number generator. One can then show that
\( \tau  \) and \( \mu  \) are obtained as \begin{equation}
\label{mathed:tweentith-eqn}
\tau =(\frac{1}{a_{0}})\ln (\frac{1}{r_{1}})
\end{equation}
and \( \mu  \) is taken to be the integer for which \begin{equation}
\label{mathed:tweentyfirst-eqn}
\sum ^{\mu -1}_{\nu =1}a_{\nu }<r_{2}a_{0}\leq \sum ^{\mu }_{\nu =1}a_{\nu }
\end{equation}

A rigorous proof of the formulae (19) - (20) is given in Ref.\cite{key-28}.
Once \( \tau  \) and \( \mu  \) are known, the time evolution of
the reacting system is specified. The stochastic rate constant \( c \)\( _{\mu } \)
is related to the more familiar deterministic reaction rate constant
\( k \)\( _{\mu } \) through simple relations. In the case of first
order reactions, both constants have the same value. In the case of
second order reactions, the rate constant is divided by the volume
of the system. We have applied the GA to our model of gene expression
involving N = 15 types of biomolecules participating in M = 16 biochemical
reactions. In the initial state (time \( t \)=0), the number of free
operator and promoter sites is 1. The number of R, I, RNAP, and ribosome
molecules is \( N_{R}= \) 20, \( N_{I}= \) 20, \( N_{RNAP}= \)
400 and \( N_{Rib}= \) 350 respectively. The number of all the other
biomolecules is set to zero at \( t \) = 0. The simulation time is
up to 2000s, i.e., less than the cell generation time typically in
the range 2000-3000s. As already mentioned, knowledge of \( \tau  \)
and \( \mu  \) enables one to calculate the numbers of biomolecules
at time \( t \)+\( \tau  \) and in this way, through repeated applications
of the GA, one can keep track of how the different numbers change
as a function of time. Fig.1 shows the number of protein molecules
present in the system as a function of time. The stochastic rate constants
of the sixteen reactions are c\( _{1} \) = 0.5, c\( _{2} \) = 0.004,
c\( _{3} \) = 0.02, c\( _{4} \) = 0.001, c\( _{5} \) = 0.8, c\( _{6} \)
= 0.9, c\( _{7} \) = 0.08, c\( _{8} \) = 0.01, c\( _{9} \) = 0.001,
c\( _{10} \) = 0.3, c\( _{11} \) = 0.8, c\( _{12} \) = 0.7, c\( _{13} \)
= 0.003, c\( _{14} \) = 0.7, c\( _{15} \) = 0.001 and c\( _{16} \)
= 0.3 respectively. The topmost curve corresponding to \( N_{I} \)
shows that the protein number reaches a steady level with fluctuations
around the mean. The dashed curve at the bottom of the figure shows
the number of proteins as a function of time when the number of inducer
molecules \( N_{I} \) (=3) is much less than \( N_{R} \). The protein
number in this case is very small, practically zero. The other curves
corresponds to values of \( N_{I} \) closer to \( N_{R}. \) As \( N_{I} \)
approaches \( N_{R} \), protein levels are intermediate between high
and low. Figure 2 shows the number of protein molecules as a function
of time in a different parameter region.

\begin{figure}[ihtp]
{\centering \resizebox*{4in}{!}{\includegraphics{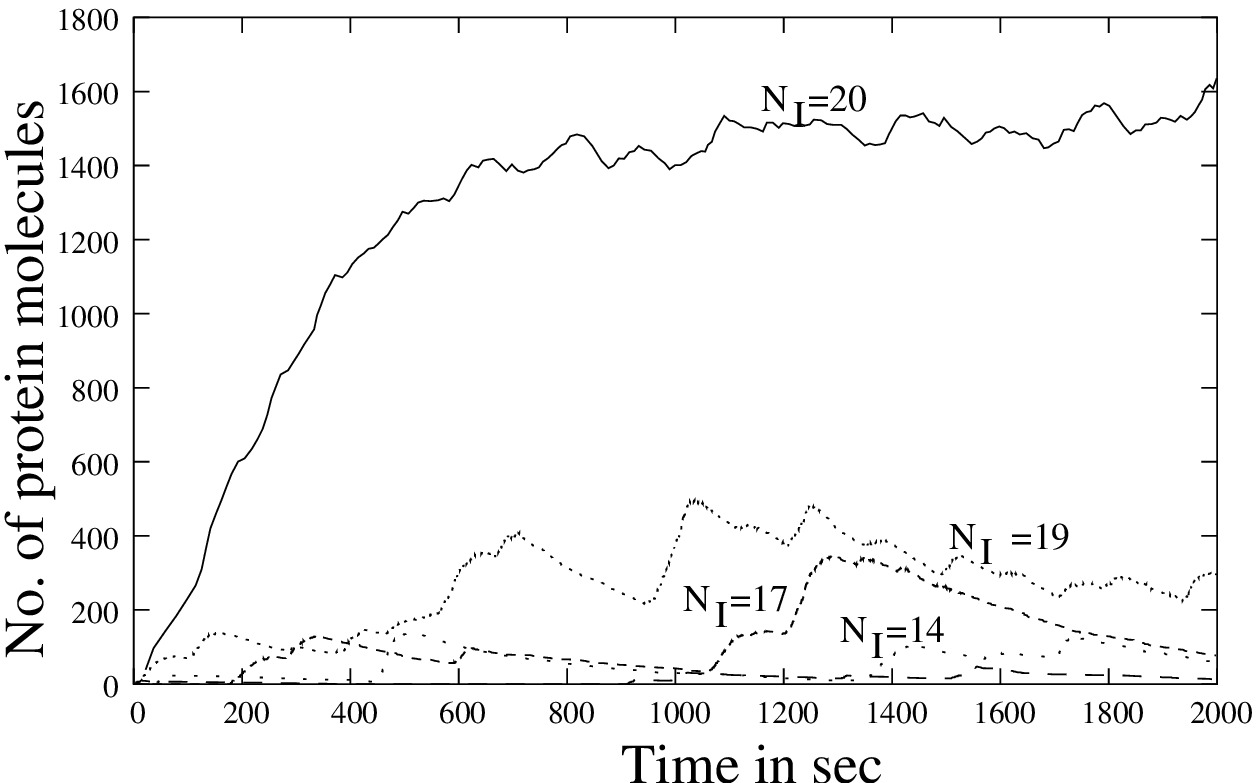}} \par}

FIG. 1. No. of protein molecules as a function of time. The stochastic
rate constants are \( c_{1}=0.5, \) \( c_{2}=0.004, \) \( c_{3}=0.02, \)
\( c_{4}=0.001, \) \( c_{5}=0.8, \) \( c_{6}=0.9, \) \( c_{7}=0.08, \)
\( c_{8}=0.01, \) \( c_{9}=0.001, \) \( c_{10}=0.3, \) \( c_{11}=0.8, \)
\( c_{12}=0.7, \) \( c_{13}=0.003, \) \( c_{14}=0.7, \) \( c_{15}=0.001, \)
\( c_{16}=0.3; \) \( N_{R}=20, \) \( N_{RNAP}=400 \) and \( N_{Rib}=350 \).
\end{figure}

\begin{figure}[ihtp]
{\centering \resizebox*{3.45in}{!}{\includegraphics{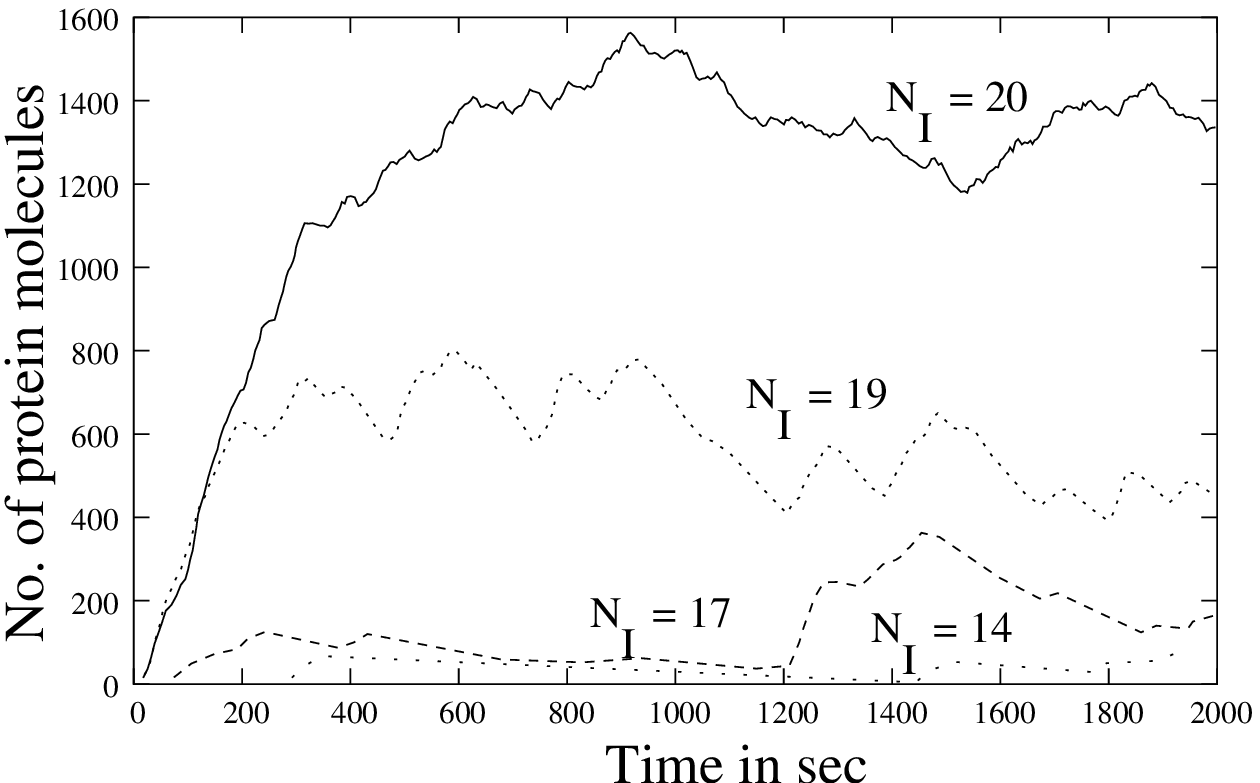}} \par}

FIG. 2. No. of protein molecules as a function of time. The stochastic
rate constants are the same as in Fig. 1 except that \( c_{1}=0.35 \)
and \( c_{14}=0.3. \) 
\end{figure}

The {}``all or none'' phenomenon has been observed in both prokaryotic
and eukaryotic cells. In prokaryotes, genes are often arranged in
operons,i.e, sets of contiguous genes which include structural and
regulatory sequences. A well-known example is that of the \emph{E.
coli} lactose (\emph{lac}) operon \cite{key-1}. Lac operon consists
of three structural genes (z, y and \( a \)) which code for the three
proteins: \( \beta  \)-galactosidase, the enzyme that catalyzes the
hydrolysis of lactose to glucose and galactose; permease, a carrier
protein responsible for membrane transport of lactose into the cell
and a third protein transacetylase. The lac operon contains three
regulatory sequences, i, P and O, which control the transcription
of mRNA leading to the synthesis of the three proteins. The sequence
i corresponds to a gene lacI which is transcribed continuously to
synthesize a repressor protein, lac repressor, at a low level. Lac
repressor binds to the operator sequence O and prevents the transcription
of the genes z, y and \( a \) so that the \( \beta  \)-galactosidase
enzyme and the permease molecules are not produced. If the bacterium
is to grow on lactose (milk sugar) which acts as its carbon source,
\( \beta  \)-galactosidase must be made available to split the sugar
into glucose and galactose. The breakdown product of lactose act as
an inducer molecule. The inducer attaches to the repressor molecule,
causing it to release the DNA so that transcription of the structural
gene is possible. The repressor is freed of the inducer when the lactose
supply is exhausted and switches off the expression of the structural
genes once more. Our simple model of gene expression incorporates
some of the key features of the lac operon. The role of the lac repressor
is played by the regulatory molecule R though its synthesis is not
explicitly considered. There is a single gene in our model analogous
to the structural gene z expressing the enzyme \( \beta  \)-galactosidase.
The inducer molecule I acts in the same manner as in the case of the
lac operon.

The experimental observation of the {}``all-or-none'' phenomenon
in the lac operon has been attributed to autocatalytic feedback mechanisms
\cite{key-22,key-23,key-27}. At low inducer concentrations, some
of the bacterial cells synthesize protein at the full rate whereas
the other cells are in the {}``off'' state. When inducer is added
to the colony of bacterial cells, simultaneous production of the \( \beta  \)-galactosidase
enzyme and permease molecules occurs. The permease molecules transport
lactose into the cell raising the internal inducer concentration which
in turn promotes the production of more \( \beta  \)-galactosidase
and permease. Thus an autocatalytic feedback process is at work and
within a short time after the appearance of the first permease molecules,
the bacterial cell becomes fully induced synthesizing the structural
proteins at maximum rate. Siegele and Hu \cite{key-24} carried out
experiments on gene expression from plasmids containing the araBAD
promoter in the presence of subsaturating concentrations of the inducer
arabinose. Again, as in the case of the lac operon, it has been suggested
that an autocatalytic induction mechanism, due to the accumulation
of inducer molecules by transport, is at work. However in all the
experiments, full induction of cells has been observed even in the
absence of autocatalysis, i.e., when the inducer availability is not
linked to that of synthesized protein molecules like permease. Our
model of gene expression does not include an autocatalytic feedback
process and a detailed analysis of the simulation results shows that
if the number of inducer \( ( \)\( \textrm{I}) \) molecules, \( N_{I} \),
is greater than or equal to the number of regulatory \( (\textrm{R}) \)
molecules, \( N_{R} \), in a cell, the cell reaches a steady state
which is state 2 (high protein level). If \( N_{I} \)= \( 0 \) or
\( \ll  \) \( N_{R} \), the cell is in state 1 (low/zero level)
but as \( N_{I} \) approaches \( N_{R} \) , protein levels intermediate
between high and low are obtained. The {}``all or none'' phenomenon
becomes more pronounced in the presence of the autocatalytic induction
(AI) mechanism. If \( N_{I} \) is originally small in a cell, the
AI mechanism leads to a rapid increase in \( N_{I} \) . When \( N_{I} \)
is \( \geq  \) \( N_{R} \), the cell is in state 2 in the steady
state. The magnitude of the protein level in state 2 is independent
of the value of \( N_{I} \) in agreement with experimental results.
Autocatalysis is responsible for a sharp bimodal distribution in protein
levels. In the absence of autocatalysis, the magnitude of the protein
level in state 2 is the same as in the case of autocatalysis but the
bimodal distribution become less sharp due to intermediate protein
levels in a fraction of cells. Our simulation results are in agreement
with experimental observations \cite{key-22,key-23,key-24}. Experiments
\cite{key-16,key-18,key-19} on single mammalian cells (eukaryotic
cells) have provided further evidence of bimodality in the distribution
of protein levels in an ensemble of cells. Again, one finds that the
amount of enhancer affects the number of expressing cells but not
the level of expression. In other words, the enhancer increases the
probability rather than the rate of transcription. In the experiments
carried out by Zlokarnik et al. \cite{key-16}, the reporter gene
synthesizes the protein \( \beta  \)-galactosidase. In unstimulared cells,
the number of these proteins is low, in the range 150 - 300. Under
the action of the stimulating agent carbachol, rapid conversion to
a state, in which 15000 - 20000 \( \beta  \)-lactonase molecules
are present in a cell, is obtained. The major conclusion of the single
cell experiments mentioned above is that in the systems considered,
the cellular state is bistable. A cell can exist in two stable steady
states: gene expression {}``off/low'' and gene expression {}``on''
with a high level of protein production. Addition of inducer/enhancer
to the system increases the fraction of cells in the {}``high''
state. In the case of eukaryotic systems, however, the mechanism of
enhancer action is not well understood. Enhancers have been suggested
to give rise to two major types of response. Enhancers increase the
rate of transcription through mainly enhancing the rate of close to
open complex formation of RNAP bound to the promoter region \cite{key-29}.
The second type of response is of the {}``all or none'' type \cite{key-25,key-56}.
Enhancers in this case do not increase the rate of transcription but
increase the fraction of cells in the high state (state 2). The {}``all
or none'' phenomenon observed in some eukaryotic systems \cite{key-16,key-18,key-19,key-25}
dose not involve autocatalysis explicitly. Thus, a more general mechanism
than in the case of prokaryotic systems is required to explain the
bimodal distributions in protein levels in an ensemble of cells.

\section*{III. Patterns of gene expression}

We now consider the model of gene expression in the absence of inducer
molecules. The values of M, the total number of reactions and N, the
number of different types of biomolecules, are both thirteen. To explore
the full parameter space, one has to treat the thirteen stochastic
rate constants corresponding to the same number of reactions as variables.
Experimental results wherever they are available, show that the usual
rate constants, \( k \)\( _{\mu } \)'s, to which the stochastic
rate constants \( c_{\mu } \)'s are related, can vary over a wide
range depending on the type of gene and the nature of the cellular
environment \cite{key-29}. Since the exploration of the full parameter
space is a daunting task, we report on some of the more general patterns
of gene expression in a restricted subspace. The effect of changing
the stochastic rate constants on specific patterns is also studied.
We have included an optional feature in our model of gene expression,
namely, cooperative binding of RNAP to the promoter region P. The
possibility of such a binding has been suggested earlier in a simplified
probabilistic model of gene expression \cite{key-40}. In the present
model, cooperative binding implies that the rate constant for the
binding of RNAP at P is enhanced by a factor q if the binding event
is immediately preceded by the Reaction 6 in which a RNAP clears the
promoter region. Cooperative binding of proteins to DNA is now well
established. In most cases of regulatory proteins, the binding cooperativity
is mediated through protein-protein interaction although increasing
evidence of DNA mediated effects have been reported \cite{key-30}.
Some recent experiments on the prokaryotic system E. coli have shown
that the transcriptional activity of the promoter is intrinsically
sensitive to the superhelical density of the DNA template \cite{key-31,key-57,key-59,key-32}.
In fact as pointed out by McClure\cite{key-29}, supercoiling gives
rise to considerably more diversity in the patterns of promoter strength
(the ability to bind weakly or strongly) than do mutations of auxiliary
proteins. There is now experimental evidence that transcription generates
increased negative supercoiling through several hundred base pairs\cite{key-31,key-57,key-59,key-32,key-33}.
In principle, this can facilitate the binding of RNAP to the DNA or
decrease the energy of activation required for the isomerization of
RNAP-promoter complex from closed to open form\cite{key-33}. Thus,
it is entirely plausible and likely that active transcription downstream
of the promoter site may lead to increased binding of RNAP (cooperative
binding) and enhanced rate of open complex formation (stochastic rate
constants \( c_{5} \)). 

We now describe three major types of gene expression pattern as a
function of time. As before, the GA is applied to the system of thirteen
biochemical reactions constituting the processes of transcription,
translation and the regulation thereof. In the gene expression pattern
designated as Type A, the protein production occurs in abrupt stochastic
bursts. A variable number of proteins is produced in each burst. The
type A pattern of gene expression has been observed experimentally\cite{key-16,key-18,key-19}
and has been attributed to stochastic effects. Figures 3 and 4 show
two such patterns of protein production for \( q=1 \) (no cooperative
binding).
\begin{figure}[ihtp]
{\centering \resizebox*{3.5in}{!}{\includegraphics{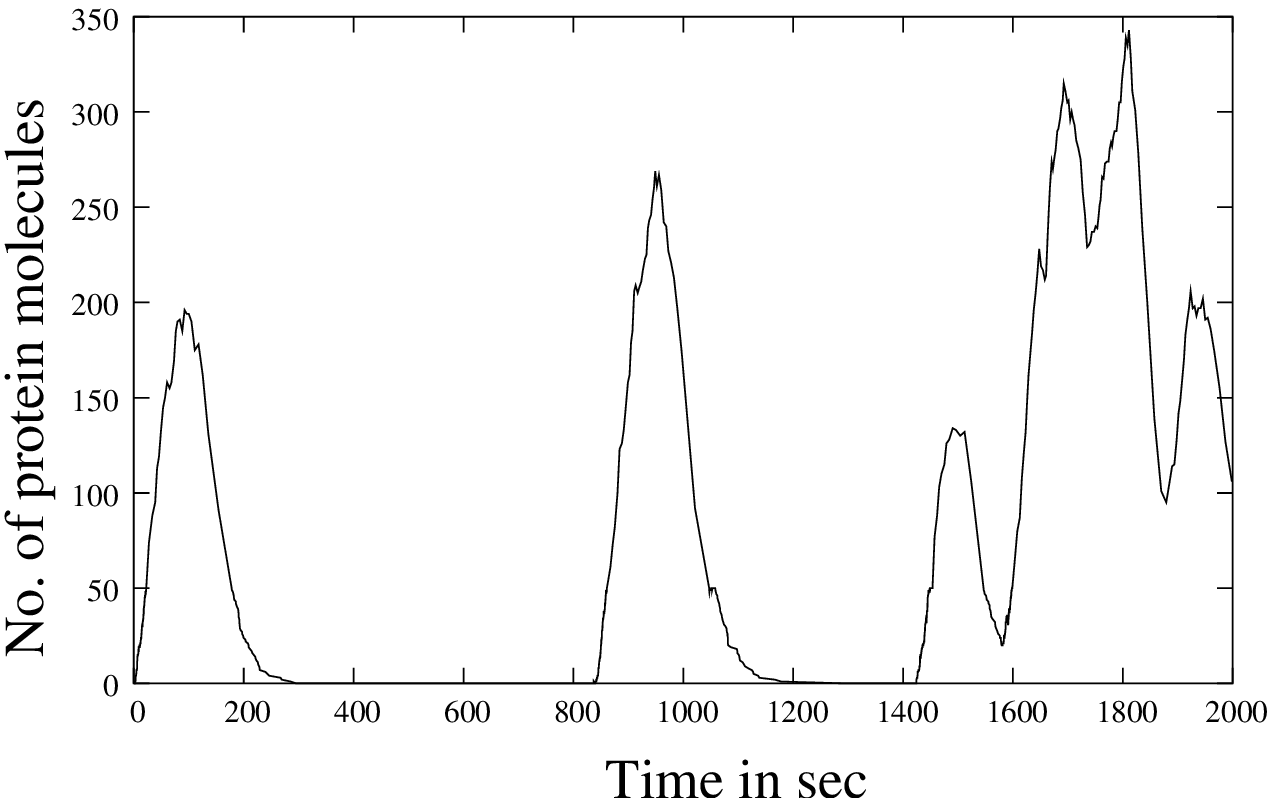}} \par}

FIG. 3. No. of protein molecules as a function of time. The stochastic
rate constants are \( c_{1}=0.008, \) \( c_{2}=0.004, \) \( c_{3}=0.007, \)
\( c_{4}=0.001, \) \( c_{5}=c_{6}=1, \) \( c_{7}=0.4, \) \( c_{8}=0.01, \)
\( c_{9}=0.001, \) \( c_{10}=0.1, \) \( c_{11}=c_{12}=1, \) \( c_{13}=0.03; \)
\( N_{RNAP}=400, \) \( N_{R}=10 \) and \( N_{Rib}=200. \)
\end{figure}

\begin{figure}[ihtp]
{\centering \resizebox*{3.5in}{!}{\includegraphics{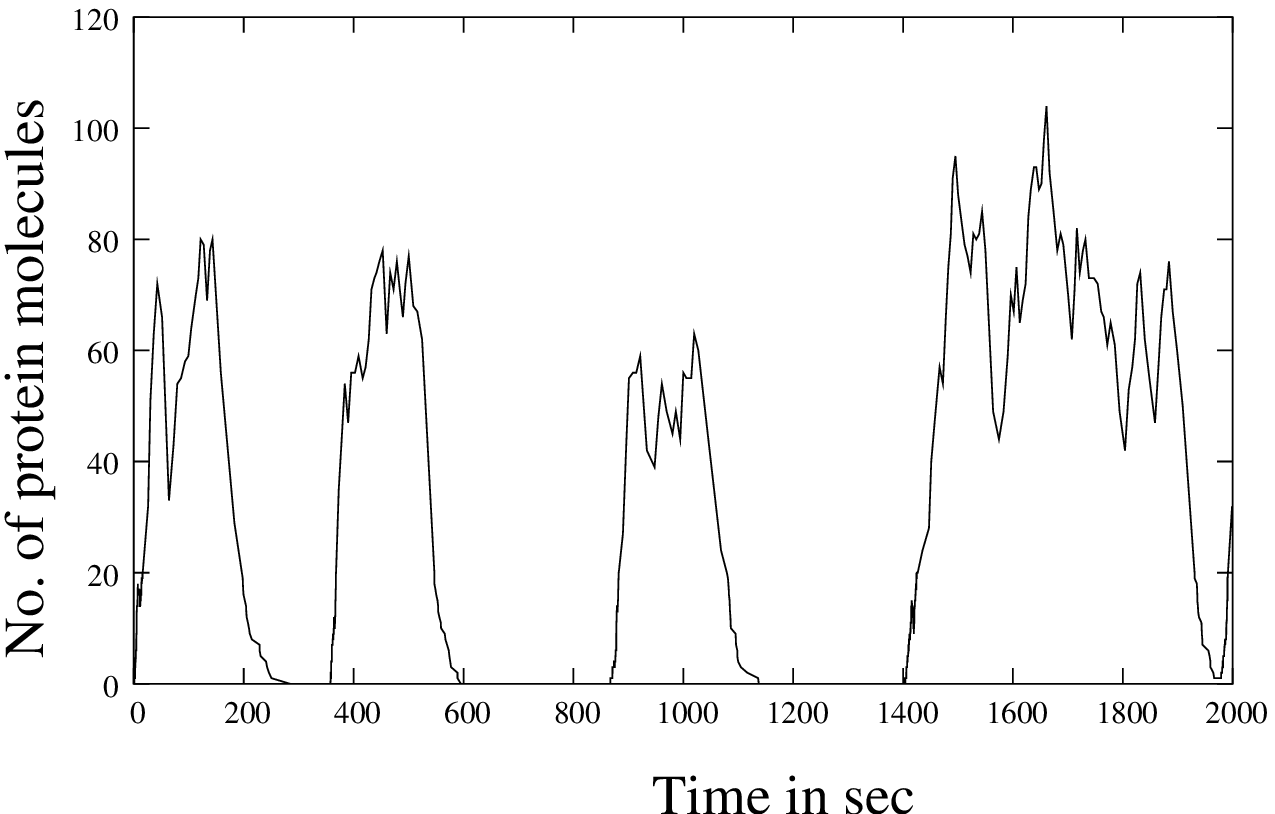}} \par}

FIG. 4. No of protein molecules as a function of time. The stochastic
rate constants are \( c_{1}=0.008, \) \( c_{2}=0.004, \) \( c_{3}=0.08, \)
\( c_{4}=0.001, \) \( c_{5}=c_{6}=1, \) \( c_{7}=0.4, \) \( c_{8}=0.01, \)
\( c_{9}=0.001, \) \( c_{10}=0.3, \) \( c_{11}=c_{12}=1, \) \( c_{13}=0.05; \)
\( N_{RNAP}=400, \) \( N_{R}=10 \) and \( N_{Rib}=200. \)
\end{figure}
 In the Type B pattern, the protein level reaches a steady state with
fluctuations around the mean (Figures 5 and 6). This type pattern
is quite common and is routinely observed in experiments.
\begin{figure}[ihtp]
{\centering \resizebox*{3.5in}{!}{\includegraphics{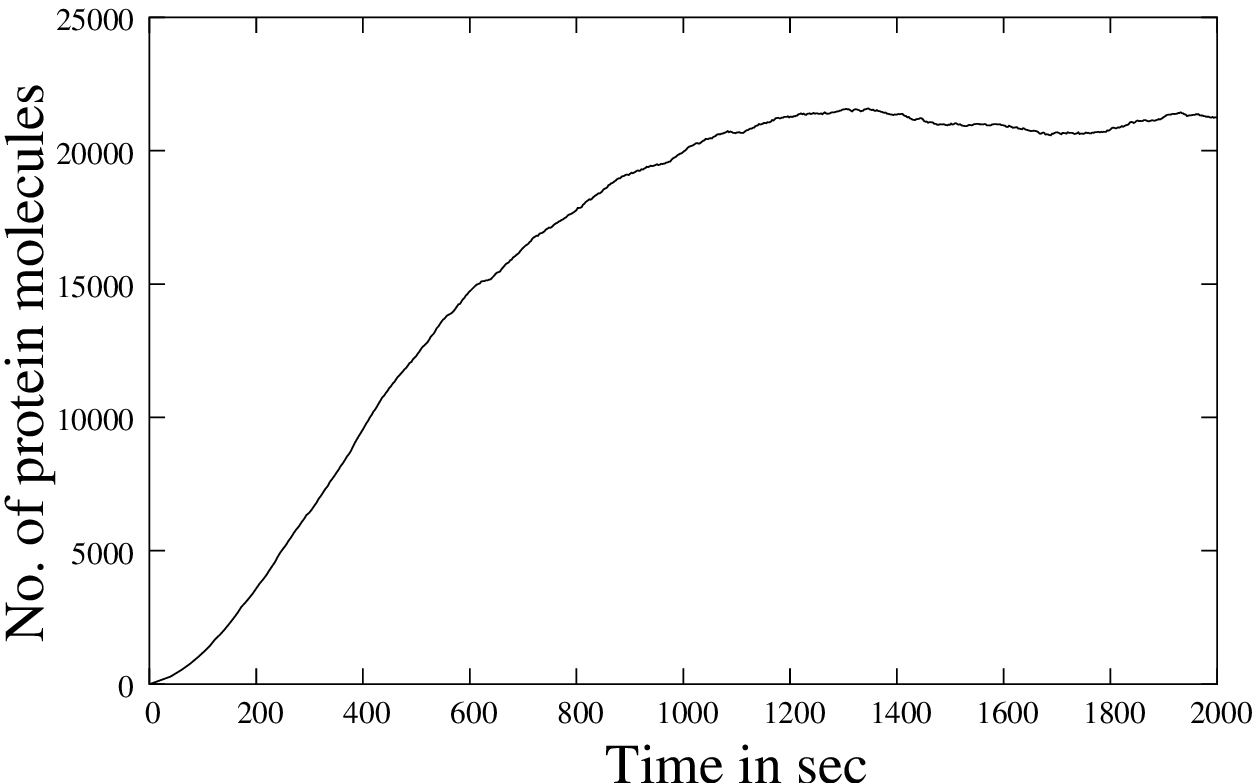}} \par}

FIG. 5. No. of protein molecules as a function of time. The stochastic
rate constants are \( c_{1}=0.008, \) \( c_{2}=0.004, \) \( c_{3}=0.5, \)
\( c_{4}=0.001, \) \( c_{5}=c_{6}=1, \) \( c_{7}=0.4, \) \( c_{8}=0.01, \)
\( c_{9}=0.001, \) \( c_{10}=0.01, \) \( c_{11}=c_{12}=1, \) \( c_{13}=0.005; \)
\( N_{RNAP}=400, \) \( N_{R}=10 \) and \( N_{Rib}=200. \) 
\end{figure}

\begin{figure}[ihtp]
{\centering \resizebox*{3.5in}{!}{\includegraphics{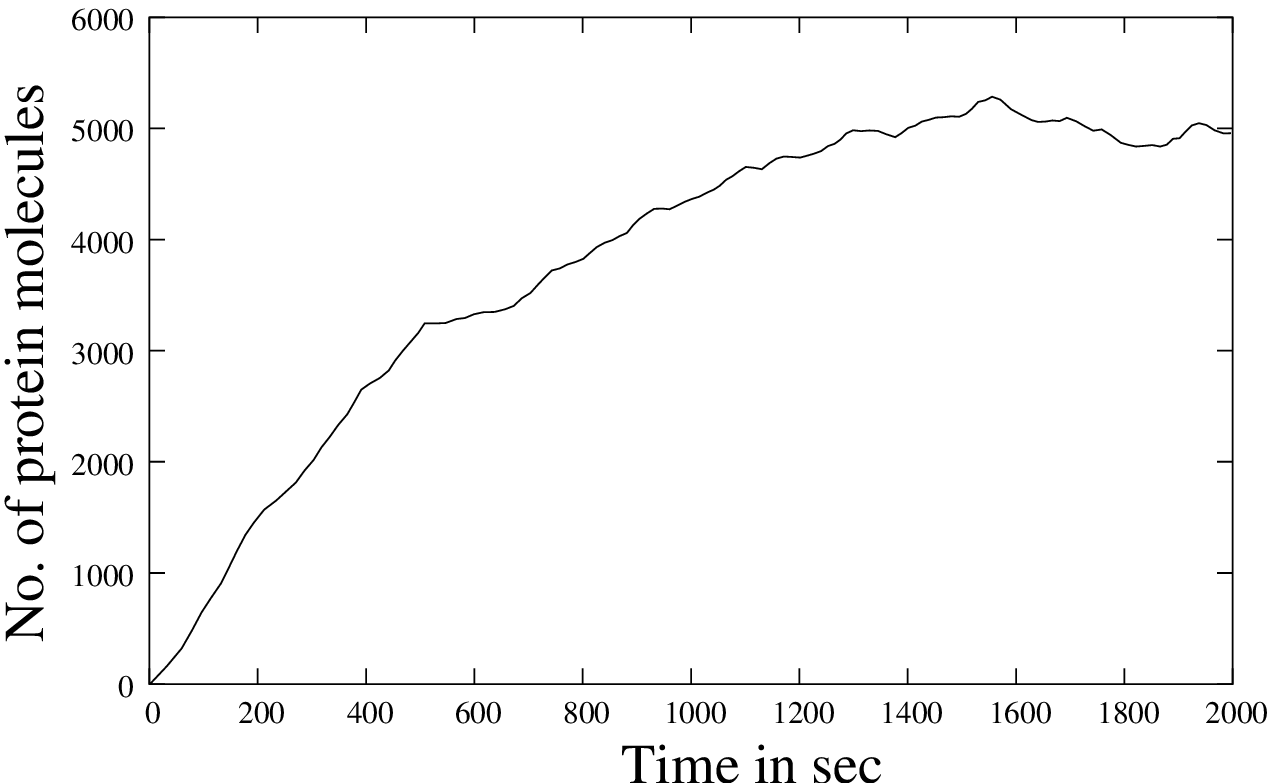}} \par}

FIG. 6. No. of protein molecules as a function of time. The stochastic
rate constants and the other parameter values are the same as in Fig.
5 except that \( c_{10}=0.1, \) and \( c_{13}=0.002. \)
\end{figure}

The Type C pattern of gene expression has an interesting structure.
As in the case of the Type A pattern, protein production occurs in
stochastic bursts, i.e., at random time intervals. The bursts may
be of various durations but in each burst, the protein number attains
the same level (with attendant fluctuations) in a very short time.
Similarly, the decay of the protein level from high to zero occurs
in a small time interval. Figures 7(a) and 7(b) show the patterns
of mRNA and protein production as a function of time. The number of
regulatory (R) molecules (\( N_{R} \)), RNAP (\( N_{RNAP} \)) and
ribosome (\( N_{Rib} \)) is 10, 400 and 200 respectively.

\begin{figure}[ihtp]
{\centering \resizebox*{2.75in}{!}{\includegraphics{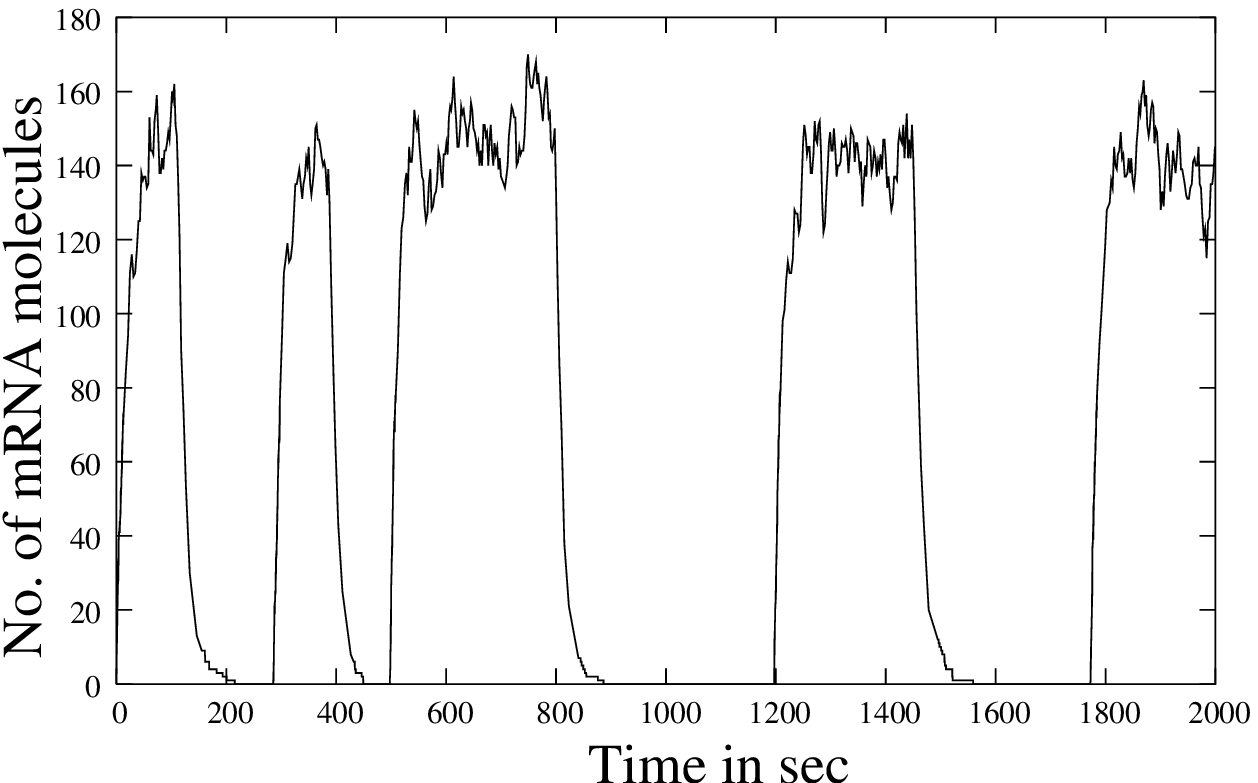}} 
\resizebox*{2.75in}{!}{\includegraphics{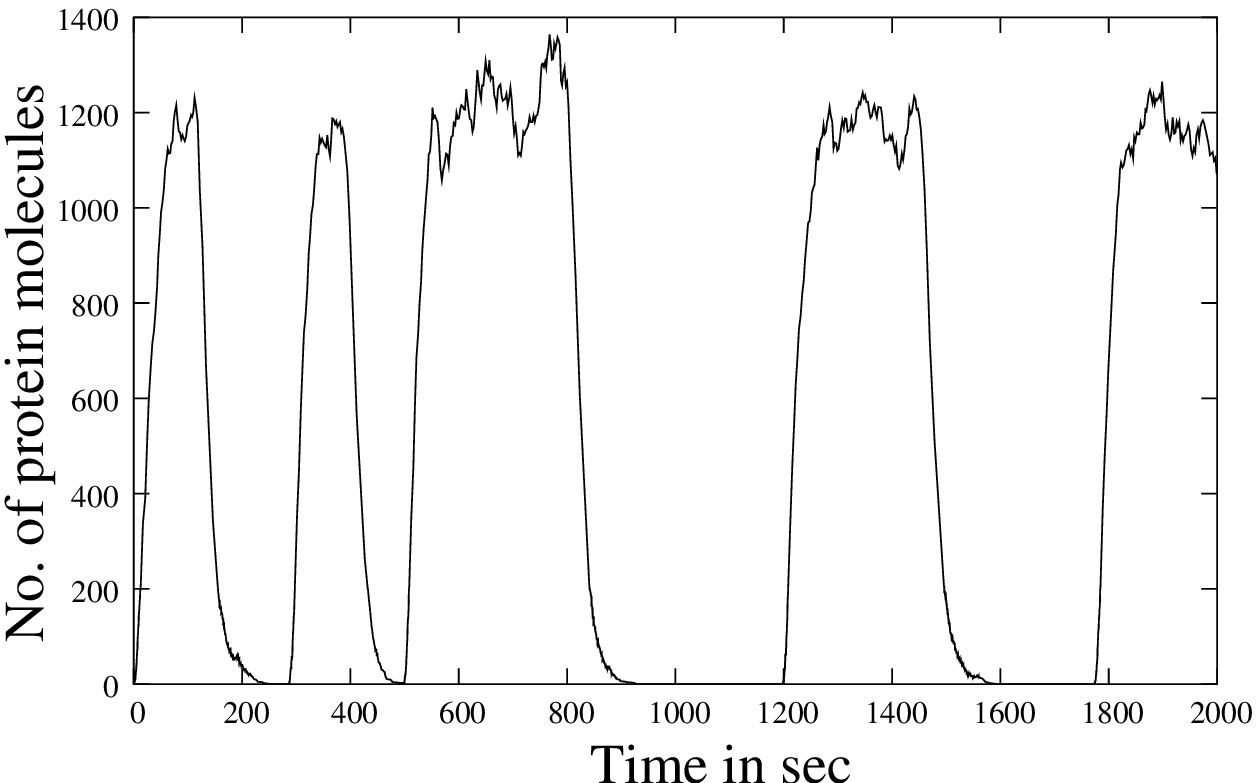}} \par}

FIG. 7. No. of mRNA (a) and protein (b) molecules as a function of
time. The stochastic rate constants are \( c_{1}=0.01, \) \( c_{2}=0.004, \)
\( c_{3}=0.7, \) \( c_{4}=0.001, \) \( c_{5}=c_{6}=20, \) \( c_{7}=0.4, \)
\( c_{8}=0.01, \) \( c_{9}=0.001, \) \( c_{10}=0.2, \) \( c_{11}=c_{12}=1, \)
\( c_{13}=0.08; \) \( N_{RNAP}=400, \) \( N_{R}=10 \) and \( N_{Rib}=200. \)
\end{figure}
 We now discuss the physical origin of the Type C pattern. The stochastic
rate constant c\( _{3} \) for RNAP binding (Reaction 3) is considerably
higher than that of the binding of R at the operator site O (Reaction
1). The initial number of RNAP molecules is also larger than that
of R molecules. As explained in section II, \( a_{\mu } \)\( dt \)
\( = \) \( h_{\mu }c_{\mu }dt \) is the probability that the \( \mu  \)
th reaction occurs in the infinitesimal time interval (\( t \) ,
\( t+dt \) ). The number of distinct molecular combinations \( h_{\mu } \)
for the \( \mu  \) th reaction is 10 and 400 for the Reactions 1
and 3 respectively. The corresponding stochastic rate constants have
the values c\( _{1} \)= 0.01 and c\( _{3} \) = 0.7. Thus, Reaction
3 is more probable than Reaction 1. Note that the stochastic rate
constants c\( _{5} \) , c\( _{6} \) are considerably high. Reactions
5 and 6 are associated with the transcription process, namely, isomerization
of the closed complex of RNAP bound to the promoter region P to the
open complex and subsequent clearance of the promoter region by RNAP.
After the binding of a RNAP to P, a host of factors including high
values of some of the relevant rate constants, leads to a sharp rise
in the number of proteins to a level determined by the transcription,
translation and protein degradation rates. The protein level is maintained
over a time interval due to the balancing of the rates of synthesis
and degradation. Binding of the R molecule to O, though less probable
than that of RNAP at P, can occur with a finite probability. Once
the R molecule is bound to O, it continues to remain bound for some
time as the dissociation rate (c\( _{2} \) = 0.004) is low. This
prevents the binding of a RNAP to P during the time interval in which
R stays bound to O, leading to a sharp fall in the number of proteins
to zero. The subsequent dissociation of the R molecule from the operator
O, followed by the binding of a RNAP to the promoter region P, tilts
the balance in favour of state 2. The cellular state thus flips from
state 1 (no. of proteins zero) to state 2 (no. of proteins high) and
vice versa at random time intervals dictated by stochastic binding
and dissociation events at O. In section II, we have discussed a bimodal
distribution in protein levels due to the {}``all or none'' phenomenon
in an ensemble of cells. In the present case, a bimodal distribution
in protein levels is obtained in an ensemble of cells if protein levels
are measured at particular instant of time. The heights and widths
of the two peaks may change as a function of time but the distribution
remains bimodal.

\begin{figure}[ihtp]
{\centering \resizebox*{3.2in}{!}{\includegraphics{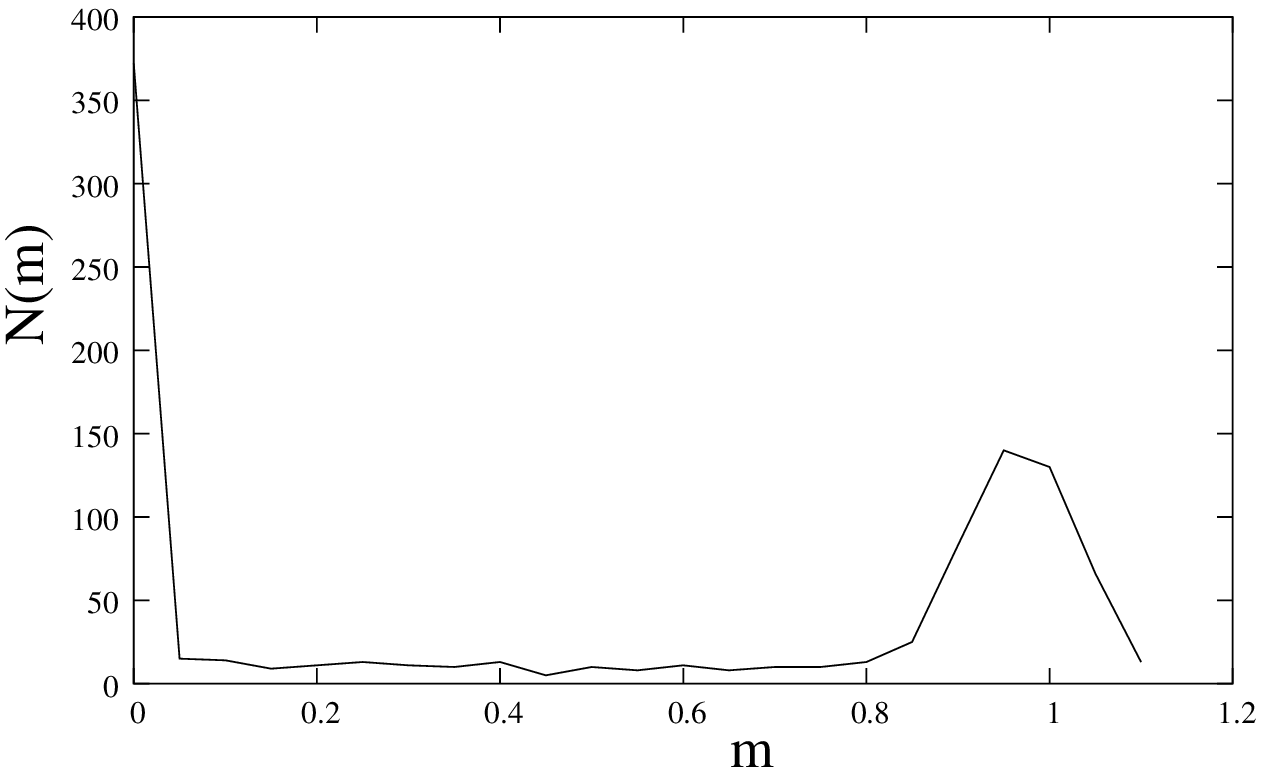}} \par}

FIG. 8. Distribution of the no. N(m) of cells expressing fraction
m of the average number of proteins. The total number of cells is
1000. The stochastic rate constants are the same as in Fig. 7.
\end{figure}

\begin{figure}[ihtp]
{\centering \subfigure{\resizebox*{3.5in}{!}{\includegraphics{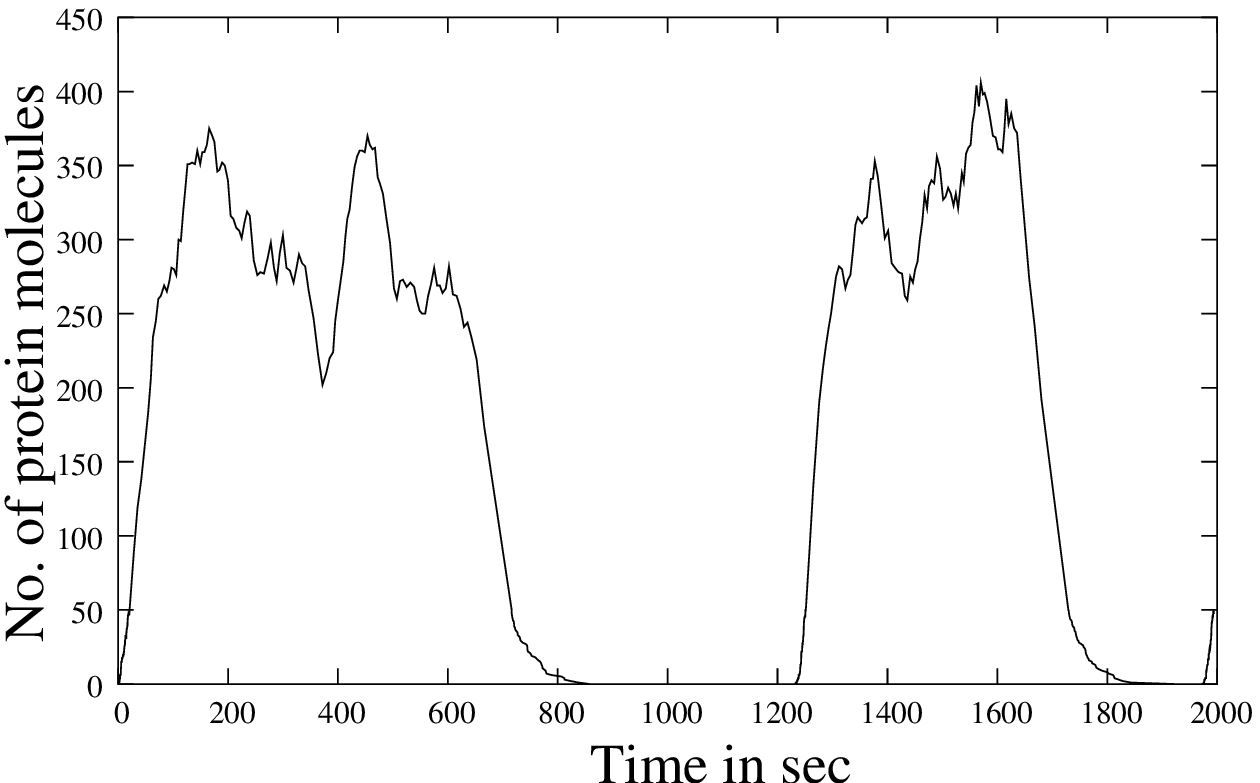}}} \par}

FIG. 9. No. of protein molecules as a function of time. The stochastic
rate constants are \( c_{1}=0.008, \) \( c_{2}=0.004, \) \( c_{3}=0.04, \)
\( c_{4}=0.001, \) \( c_{5}=c_{6}=1, \) \( c_{7}=0.4, \) \( c_{8}=0.01, \)
\( c_{9}=0.001, \) \( c_{10}=0.1, \) \( c_{11}=c_{12}=1, \) \( c_{13}=0.03; \)
\( N_{RNAP}=400, \) \( N_{R}=10 \) and \( N_{Rib}=200. \)
\end{figure}

Figure 8 shows the distribution of the number N(m) of cells expressing
a fraction m of the average number of proteins at a particular instant
of time. The total number of cells is 1000. The values of the stochastic
rate constants are the same as in Figs. 7(a) and 7(b). Type C pattern
of gene expression is also obtained for lower values of \( c_{5} \)
and \( c_{6} \) (Fig. 9) though better quality patterns are obtained
for high values of the rate constants. Figures 10(a) and 10(b) show
the temporal variations of the mRNA and protein numbers with stochastic
rate constants the same as in Figs. 7(a) and 7(b) but the enhancement
factor q, associated with cooperative RNAP binding, has been raised
from 1 to 10. Comparing the two sets of Figures, one concludes that
cooperativity increases the duration of state 2 ({}``high'' level).
The magnitude of the mean level remains unchanged. In an ensemble
of cells, a greater fraction of cells is in state 2 than in the earlier
case. The origin and nature of bimodal distribution in protein levels
are different for the model system considered in section II and the
type C pattern of gene expression. In the latter case, no autocatalytic
feedback is necessary to obtain a bimodal distribution. Positive (autocatalytic)
feedback mechanism has been invoked to explain the {}``all-or-none''
phenomenon in prokaryotic\cite{key-22,key-23,key-24} and eukaryotic\cite{key-56}systems.
Experimental reports of the phenomenon in some eukaryotic systems\cite{key-16,key-18,key-19}
suggest that autocatalytic (positive) feedback is not essential for
the obsevance of the phenomenon. In these systems, activation to the
high protein level is enhancer mediated. The origin of bimodal distribution
in protein levels is, however, yet to be elucidated. 
\begin{figure}[ihtp]
{\centering \resizebox*{2.75in}{!}{\includegraphics{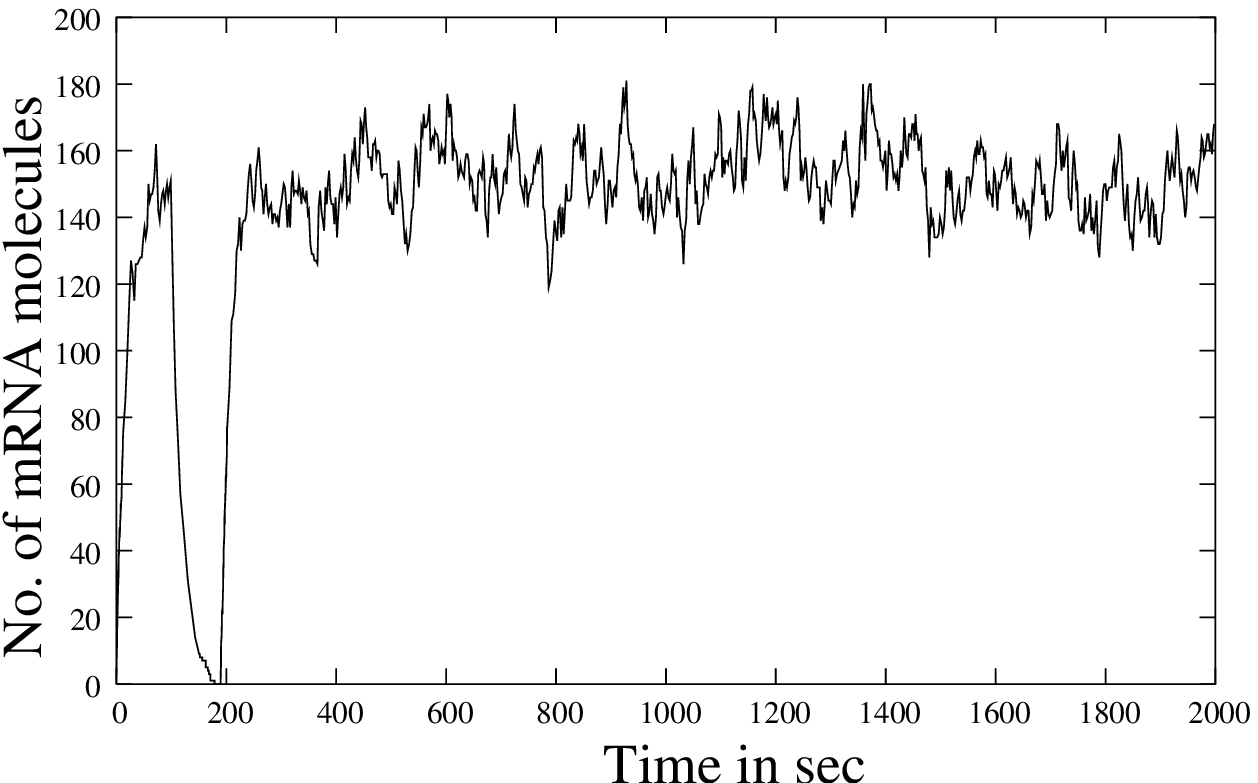}} 
\resizebox*{2.75in}{!}{\includegraphics{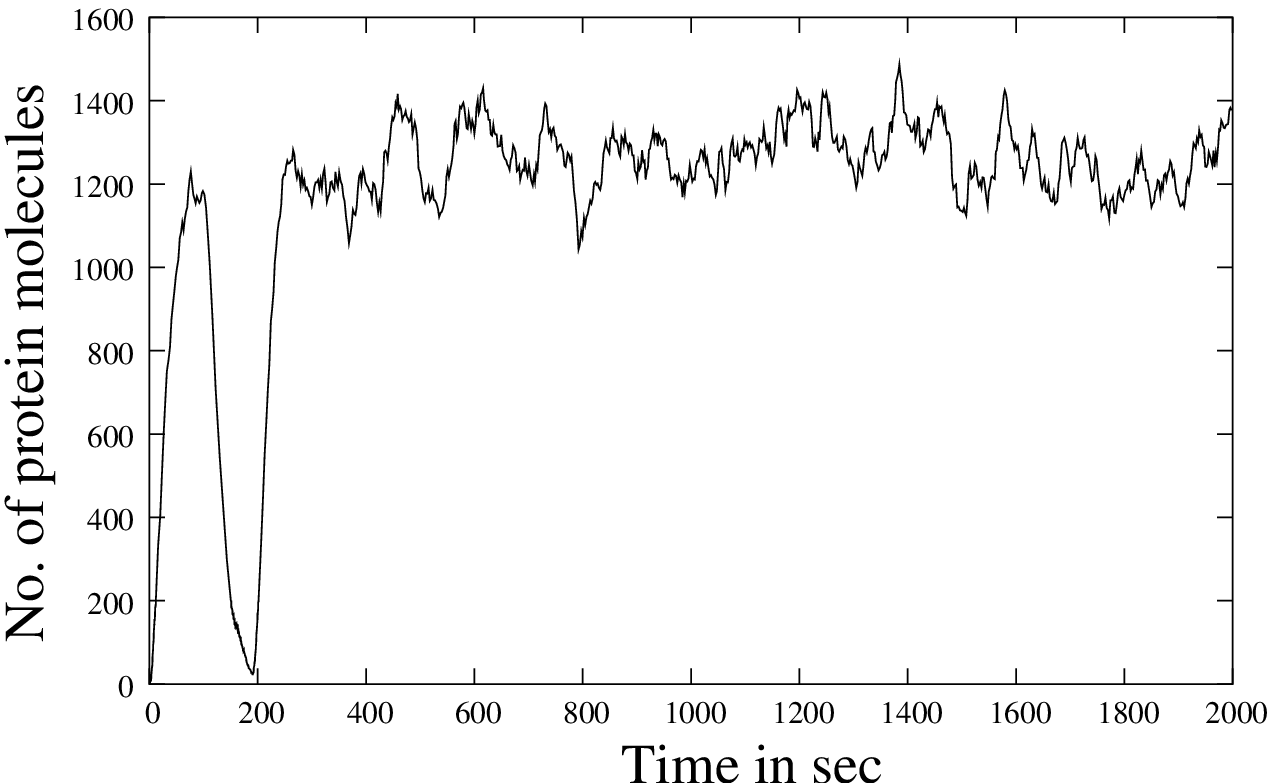}} \par}

FIG. 10. No. of mRNA (a) and protein (b) molecules as a function of
time. The stochastic rate constants are \( c_{1}=0.01, \) \( c_{2}=0.004, \)
\( c_{3}=0.7, \) \( c_{4}=0.001, \) \( c_{5}=c_{6}=20, \) \( c_{7}=0.4, \)
\( c_{8}=0.01, \) \( c_{9}=0.001, \) \( c_{10}=0.2, \) \( c_{11}=c_{12}=1, \)
\( c_{13}=0.08; \)  \( N_{RNAP}=400, \) \( N_{R}=10, \) \( N_{Rib}=200 \)
and \( q=10. \) 
\end{figure}
 Figure 11 shows the number of proteins (solid line) and mRNA molecules
(dotted line) as a function of time to make a simultaneous comparison
of the rise and decay of protein and mRNA levels.Note that after the
number of mRNA molecules become zero, there is a time delay before
the number of proteins falls to zero. Even when there are no mRNA
molecules in the system, some proteins remain which degrade to zero
level at a decay rate lower than that of the mRNA molecules. In Fig.
12, a pattern of gene expression is shown in which the protein number
never falls to zero and a variable number of proteins is synthesized
as a function of time.

We have further checked how robust pattern C ( the gene expression
pattern shown in Figure 7) is when the different stochastic parameters
are changed. 
\begin{figure}[ihtp]
{\centering \resizebox*{3.5in}{!}{\includegraphics{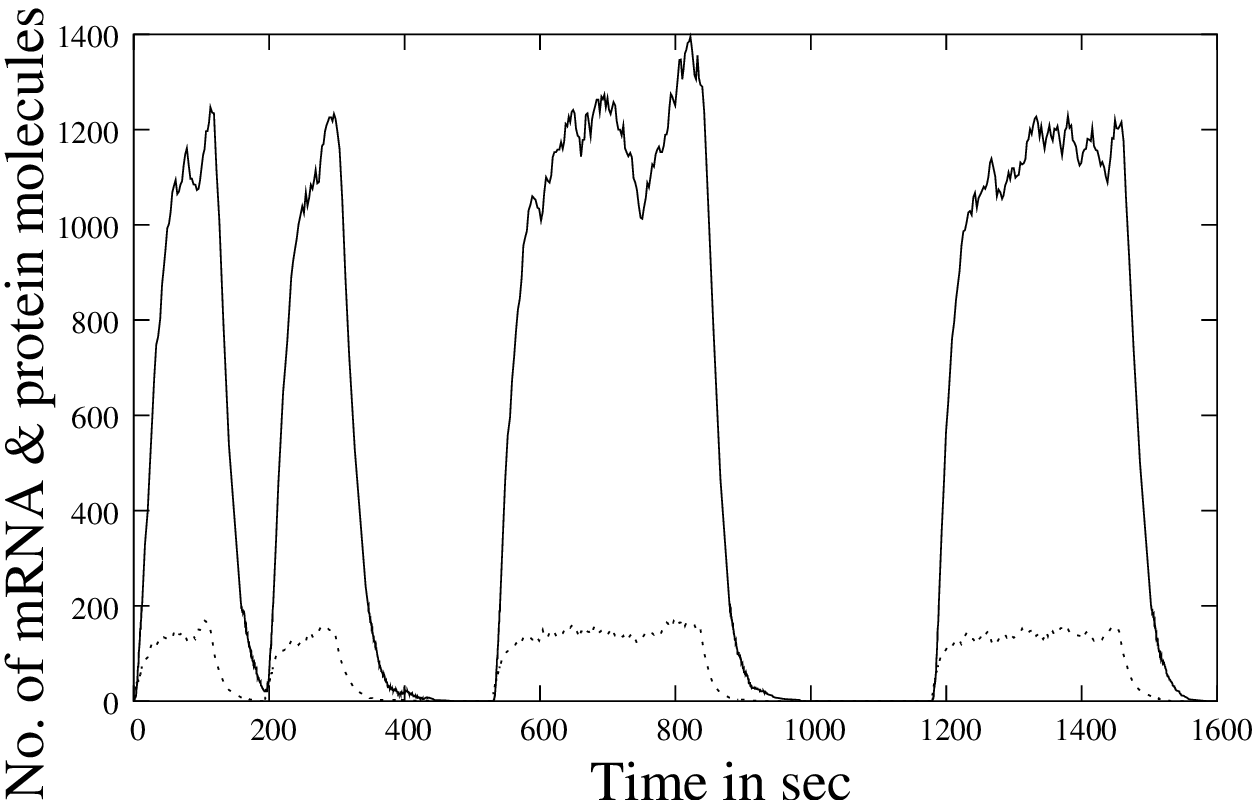}} \par}

FIG. 11. No. of protein and mRNA molecules as a function of time.
The stochastic rate constants are \( c_{1}=0.01, \) \( c_{2}=0.004, \)
\( c_{3}=0.5, \) \( c_{4}=0.001, \) \( c_{5}=c_{6}=20, \) \( c_{7}=0.4, \)
\( c_{8}=0.01, \) \( c_{9}=0.001, \) \( c_{10}=0.2, \) \( c_{11}=c_{12}=1, \)
\( c_{13}=0.08; \) \( N_{RNAP}=400, \) \( N_{R}=10, \) \( N_{Rib}=200. \)
\end{figure}

\begin{figure}[ihtp]
{\centering \resizebox*{3.5in}{!}{\includegraphics{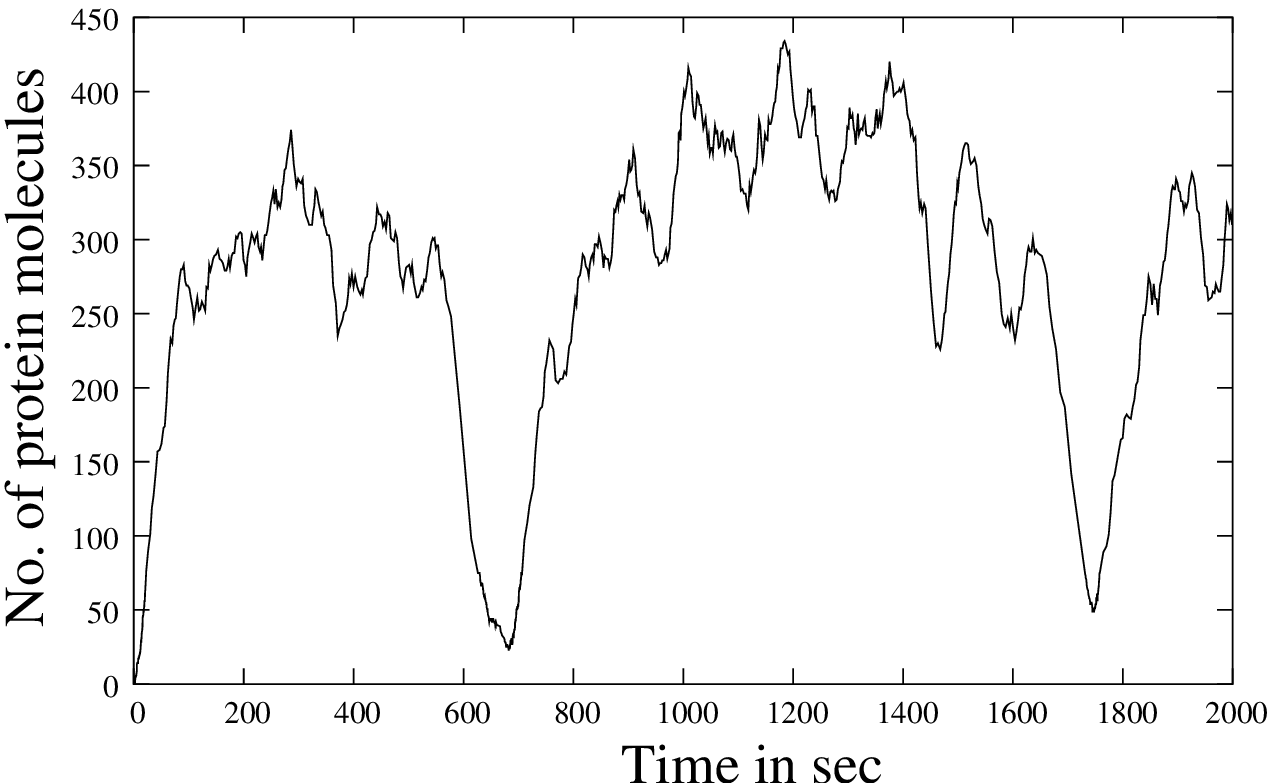}} \par}

FIG. 12. No. of protein molecules as a function of time. The stochastic
rate constants are \( c_{1}=0.008, \) \( c_{2}=0.004, \) \( c_{3}=0.08, \)
\( c_{4}=0.001, \) \( c_{5}=c_{6}=1, \) \( c_{7}=0.4, \) \( c_{8}=0.01, \)
\( c_{9}=0.001, \) \( c_{10}=0.1, \) \( c_{11}=c_{12}=1, \) \( c_{13}=0.03; \)
\( N_{RNAP}=400, \) \( N_{R}=10, \) \( N_{Rib}=200. \)
\end{figure}

We change one parameter at a time keeping all the other parameter
values the same as in Fig. 7. As c\( _{1} \) decreases from 0.01
(c\( _{1} \) = 0.01 in Fig. 7), the duration of state 2 increases
and ultimately state 2 becomes the steady state (Figs. 13(a) and 13(b)),
i.e., a Type B pattern is obtained. As c\( _{1} \) increases from
0.01, the Type C pattern is still obtained (Fig. 13(c), c\( _{1} \)
= 0.02) but for higher c\( _{1} \) values, say, c\( _{1} \) = 0.03,
the pattern of gene expression becomes of Type A (Fig. 13(d)). If
c\( _{2} \) is varied, then as c\( _{2} \) decreases from 0.004
(Fig. 7), the total duration of state 2, for \( t \) in the range
0 - 2000s, decreases (Fig. 13(e) with c\( _{2} \) = 0.008). If c\( _{4} \)
is varied, one finds that the Type C pattern of gene expression is
obtained over a wide range of values. The greater the values of c\( _{5} \)
and c\( _{6} \) , the higher is the level of proteins attained in
state 2. The best Type C patterns are obtained if the values of c\( _{5} \)
and c\( _{6} \) are high. Figs. 13(f) and 13(g) correspond to c\( _{5} \)=
2, c\( _{6} \) = 20 and c\( _{5} \)= 20, c\( _{6} \) = 2 respectively.
If c\( _{7} \) is varied, the Type C pattern is obtained over a wide
range of values. If c\( _{8} \) increases from the value 0.01 (Fig.
7), Type C pattern becomes Type A (Fig. 13(h), c\( _{8} \) = 0.015).
If c\( _{8} \) decreases from the value 0.01, the total duration
of state 2 increases(Fig. 13(i)). If c\( _{10} \) decreases from
the value 0.2 (Fig. 7), the Type C pattern is lost (Fig. 13(j), c\( _{10} \)
= 0.1). If c\( _{10} \) increases from 0.2 the Type C pattern is
obtained over a wide range of values of c\( _{10} \). The best type
of Type C pattern is obtained for similar values of c\( _{11} \)
and c\( _{12} \). 

\twocolumn
\begin{figure}[ihtp]
{\centering \resizebox*{2.85in}{!}{\includegraphics{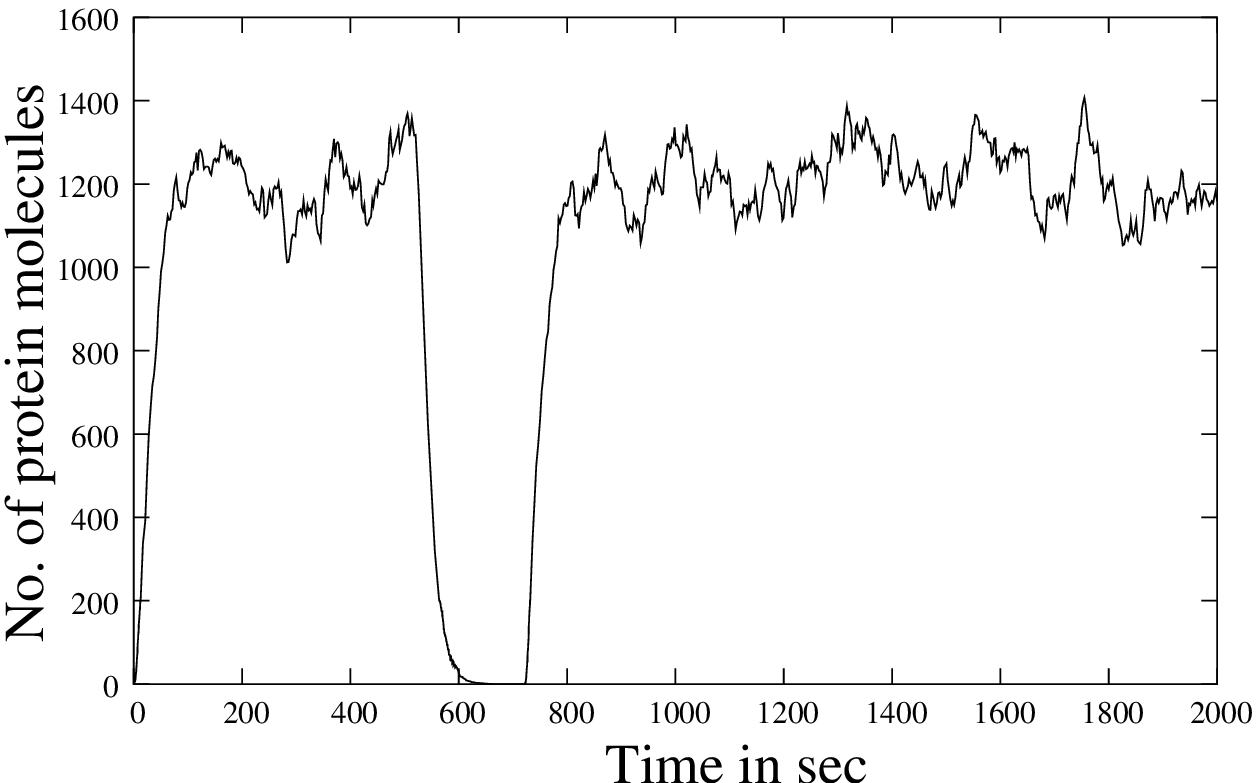}} \par}

FIG. 13(a). No. of protein molecules as a function of time. The stochastic
rate constants are \( c_{1}=0.002, \) \( c_{2}=0.004, \) \( c_{3}=0.7, \)
\( c_{4}=0.001, \) \( c_{5}=c_{6}=20, \) \( c_{7}=0.4, \) \( c_{8}=0.01, \)
\( c_{9}=0.001, \) \( c_{10}=0.2, \) \( c_{11}=c_{12}=1, \) \( c_{13}=0.08; \)
\( N_{RNAP}=400, \) \( N_{R}=10 \) and \( N_{Rib}=200. \) The stochastic
rate constants and the other parameter values are the same as in Fig.
7 except that \( c_{1}=0.002. \)
\end{figure}

\begin{figure}[ihtp]
{\centering \resizebox*{2.85in}{!}{\includegraphics{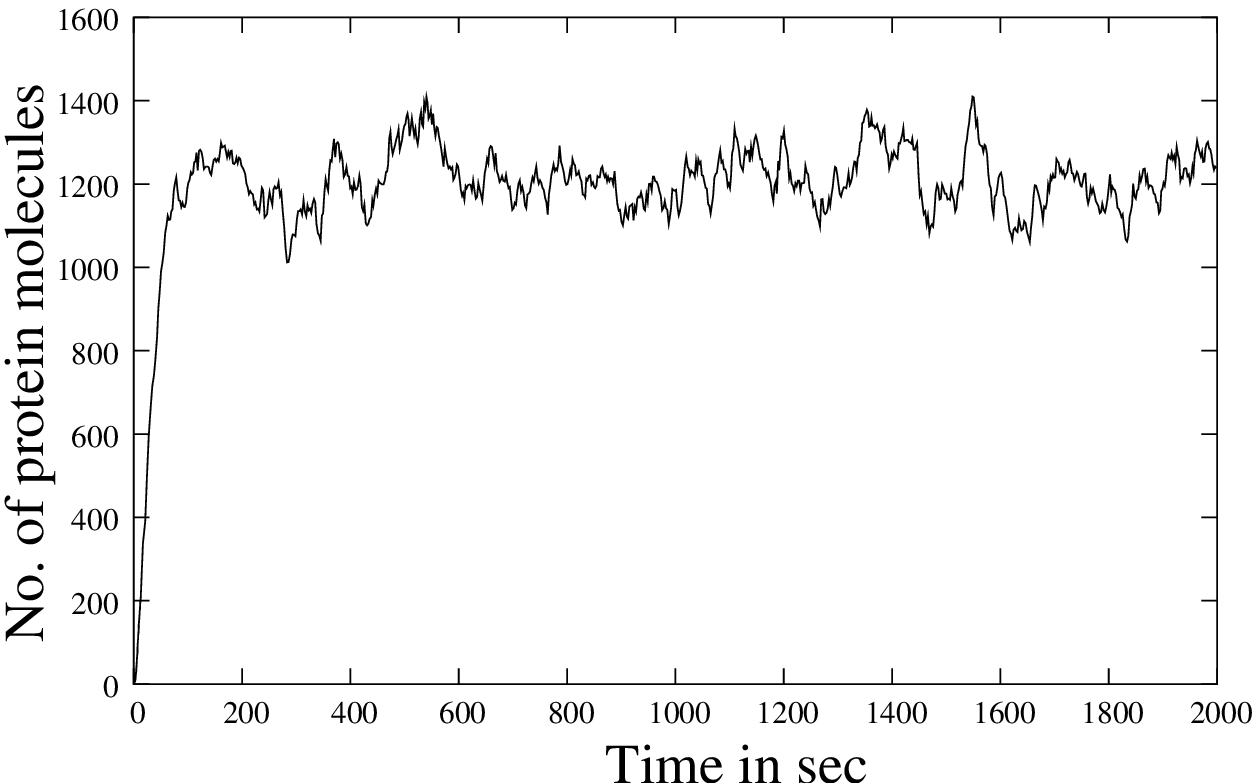}} \par}

FIG. 13(b). No. of protein molecules as a function of time. The stochastic
rate constants and the other parameter values are the same as in Fig.
13(a) except that \( c_{1}=0.0008. \) 
\end{figure}

\begin{figure}[ihtp]
{\centering \resizebox*{2.85in}{!}{\includegraphics{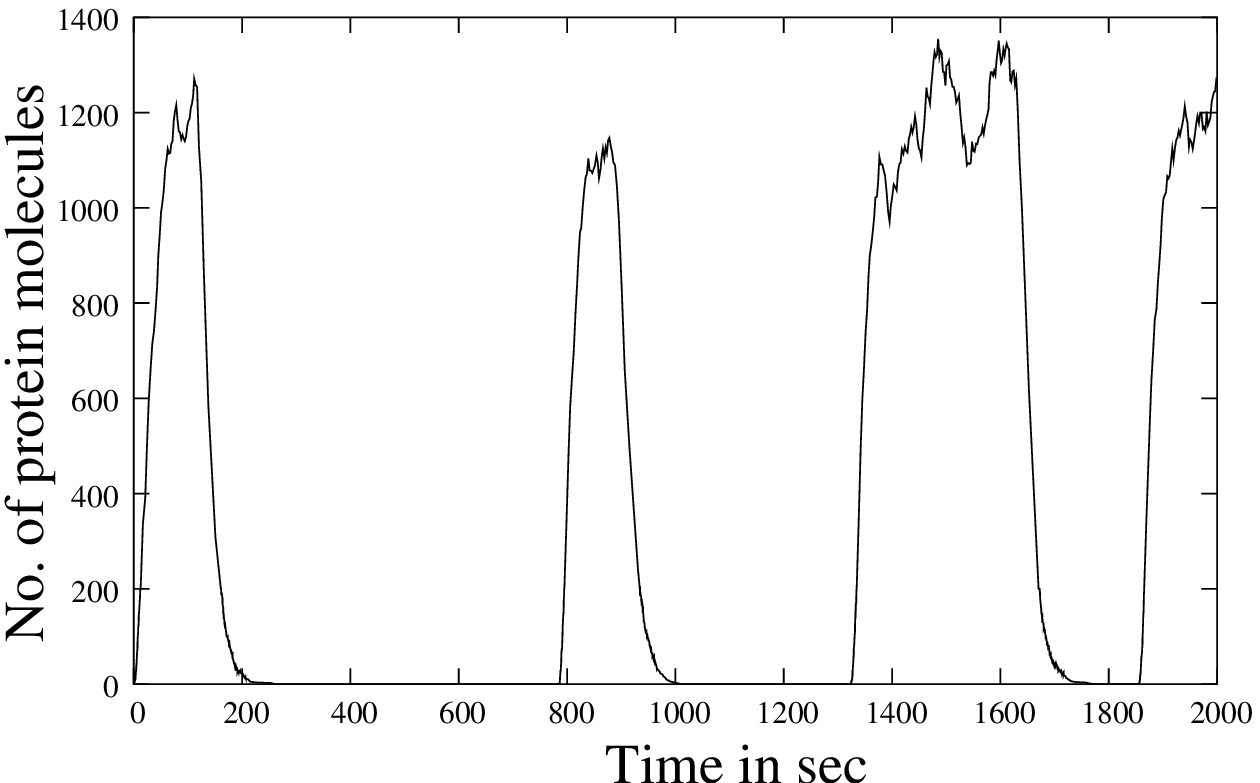}} \par}

FIG. 13(c). No. of protein molecules as a function of time. The stochastic
rate constants and the other parameter values are the same as in Fig.
13(a) except that \( c_{1}=0.02. \) 
\end{figure}

\begin{figure}[ihtp]
{\centering \resizebox*{2.85in}{!}{\includegraphics{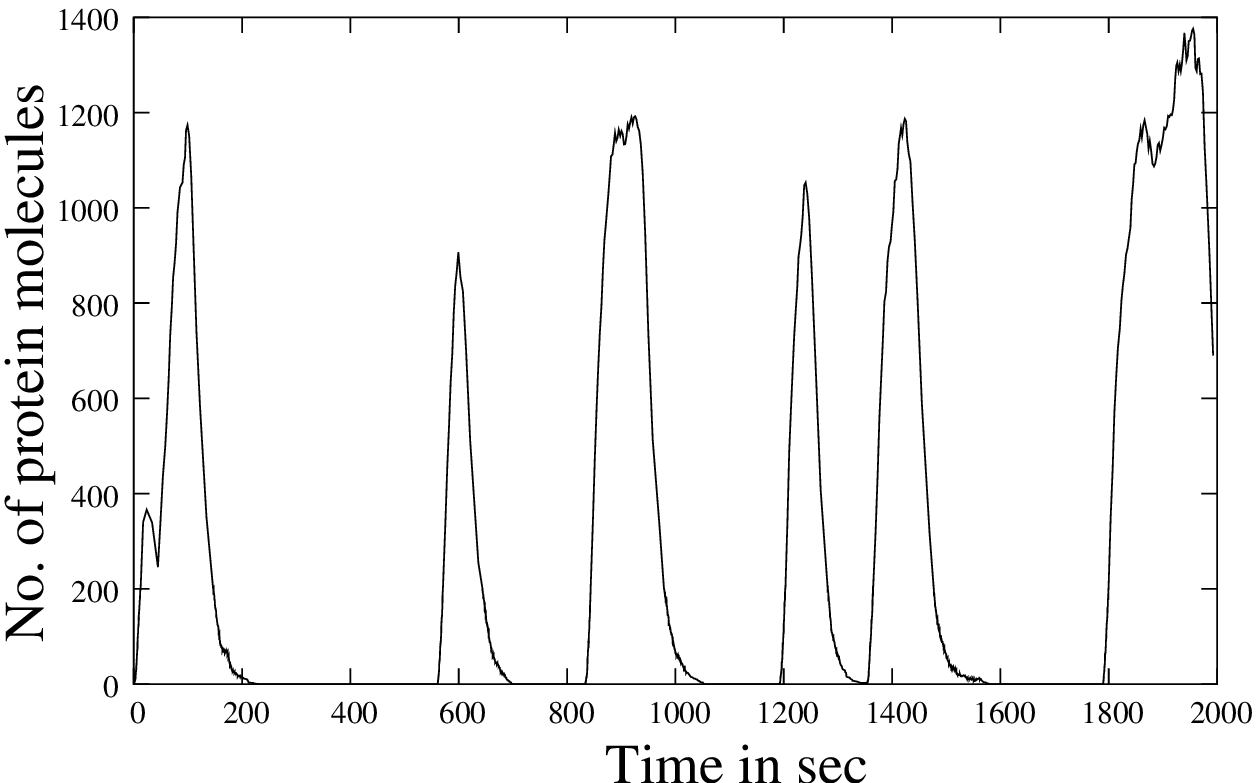}} \par}

FIG. 13(d). No. of protein molecules as a function of time. The stochastic
rate constants and the other parameter values are the same as in Fig.
13(a) except that \( c_{1}=0.03 \). 
\end{figure}

\begin{figure}[ihtp]
{\centering \resizebox*{2.85in}{!}{\includegraphics{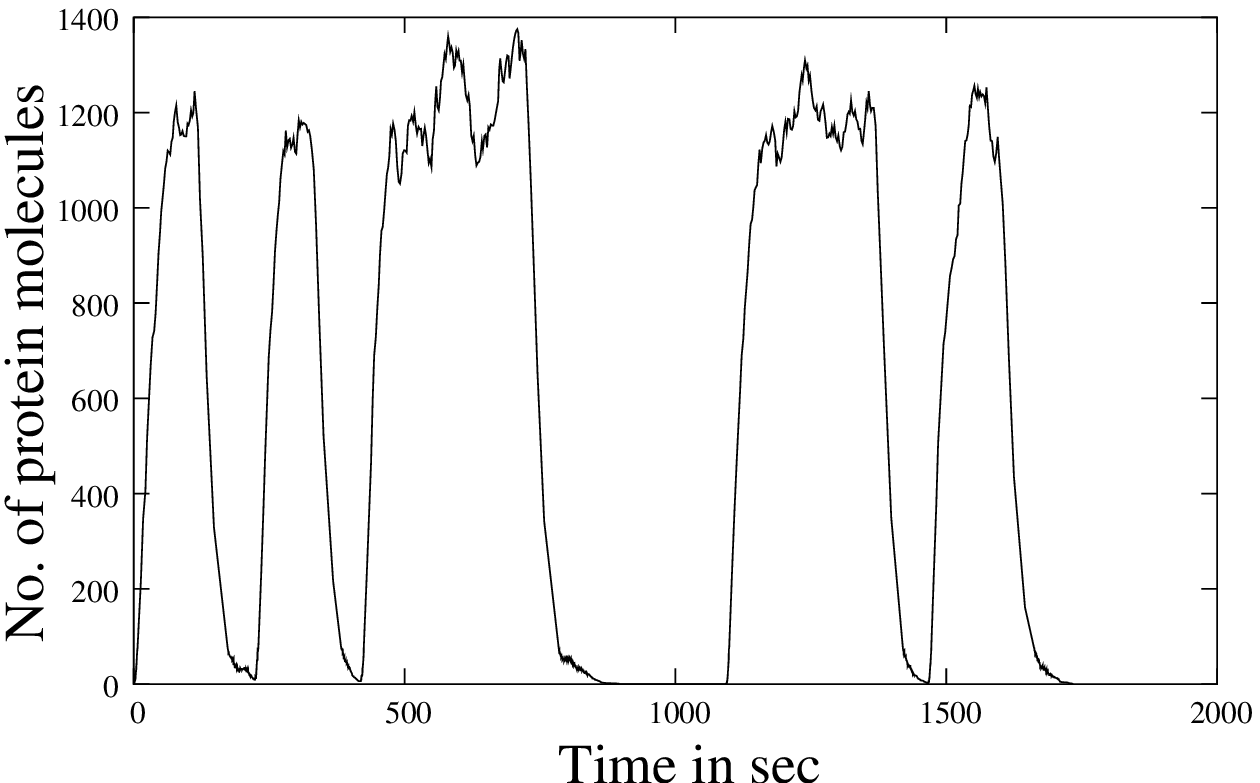}} \par}

FIG. 13(e). No. of protein molecules as a function of time. The stochastic
rate constants and the other parameter values are the same as in Fig.
7 except that \( c_{2}=0.008. \) 
\end{figure}

\begin{figure}[ihtp]
{\centering \resizebox*{2.85in}{!}{\includegraphics{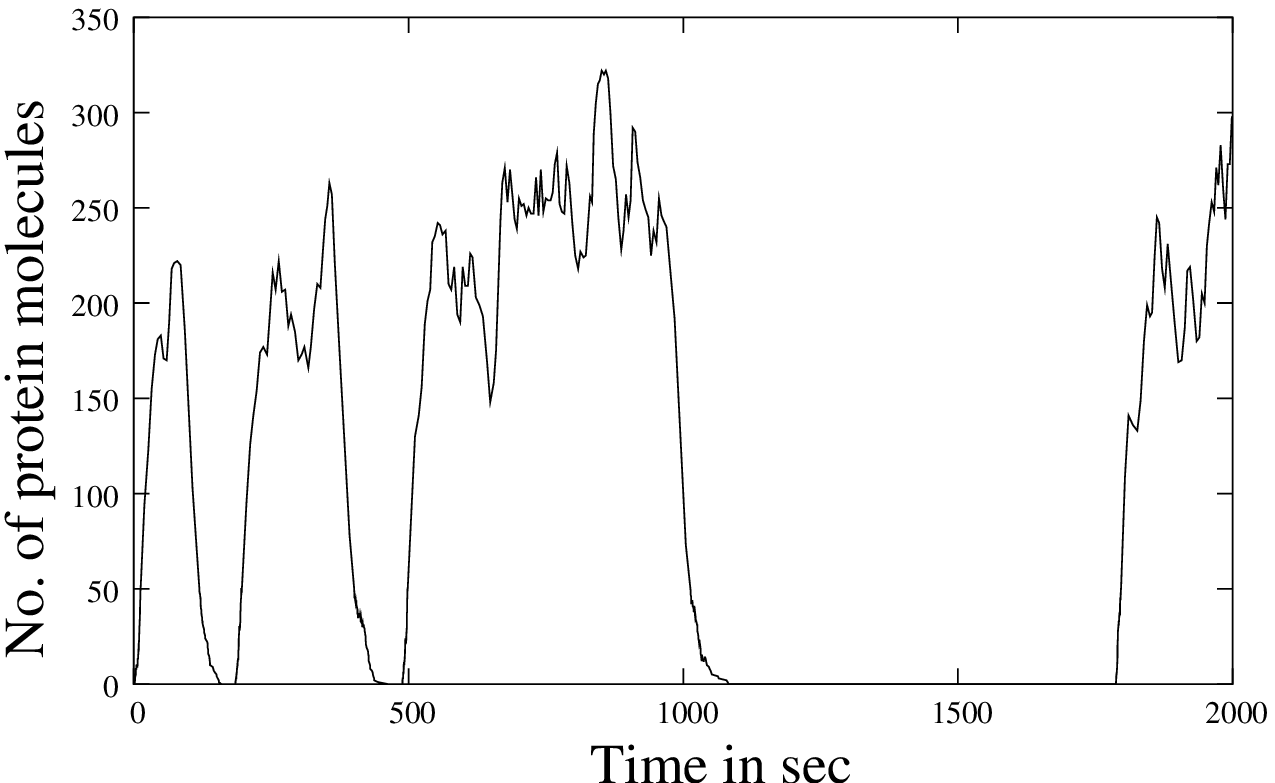}} \par}

FIG. 13(f). No. of protein molecules as a function of time. The stochastic
rate constants and the other parameter values are the same as in Fig.
7 except that \( c_{5}=2 \) and \( c_{6}=20. \) 
\end{figure}

\begin{figure}[ihtp]
{\centering \resizebox*{2.85in}{!}{\includegraphics{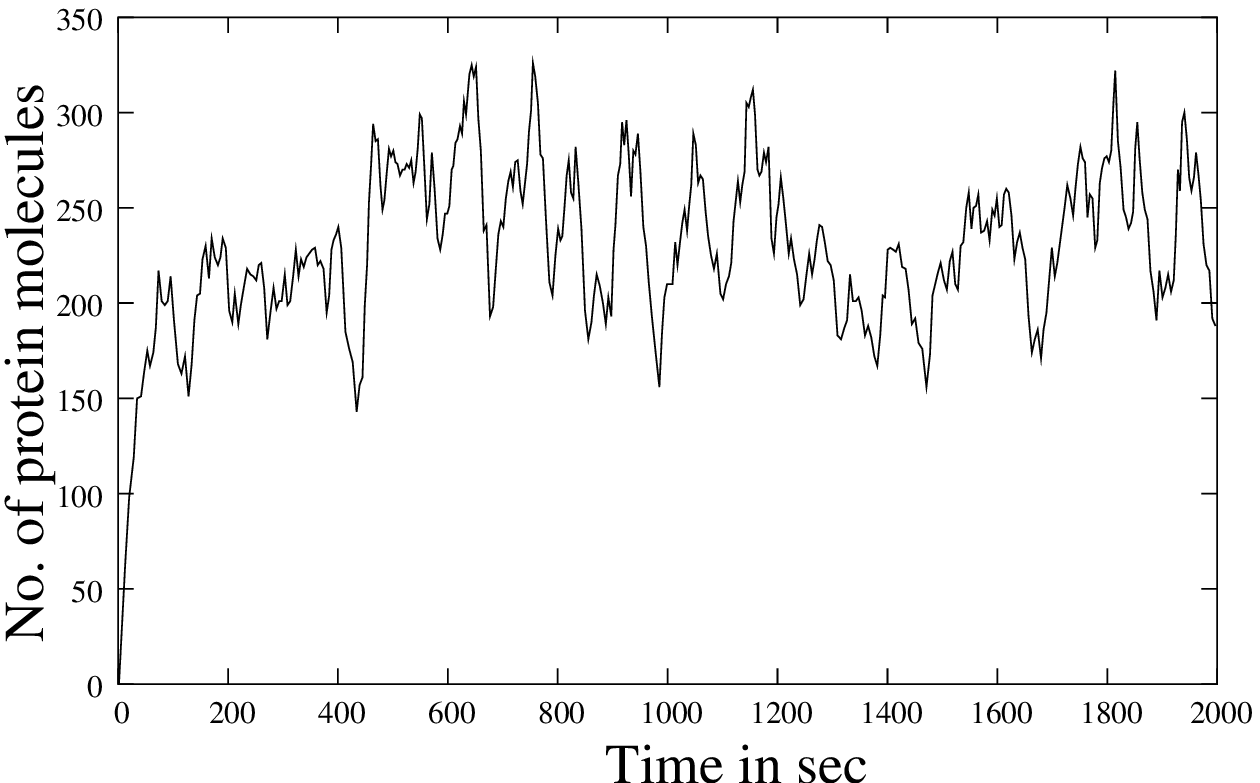}} \par}

FIG. 13(g). No. of protein molecules as a function of time. The stochastic
rate constants and the other parameter values are the same as in Fig.
13(f) except that \( c_{5}=20 \) and \( c_{6}=2. \) 
\end{figure}

\begin{figure}[ihtp]
{\centering \resizebox*{2.85in}{!}{\includegraphics{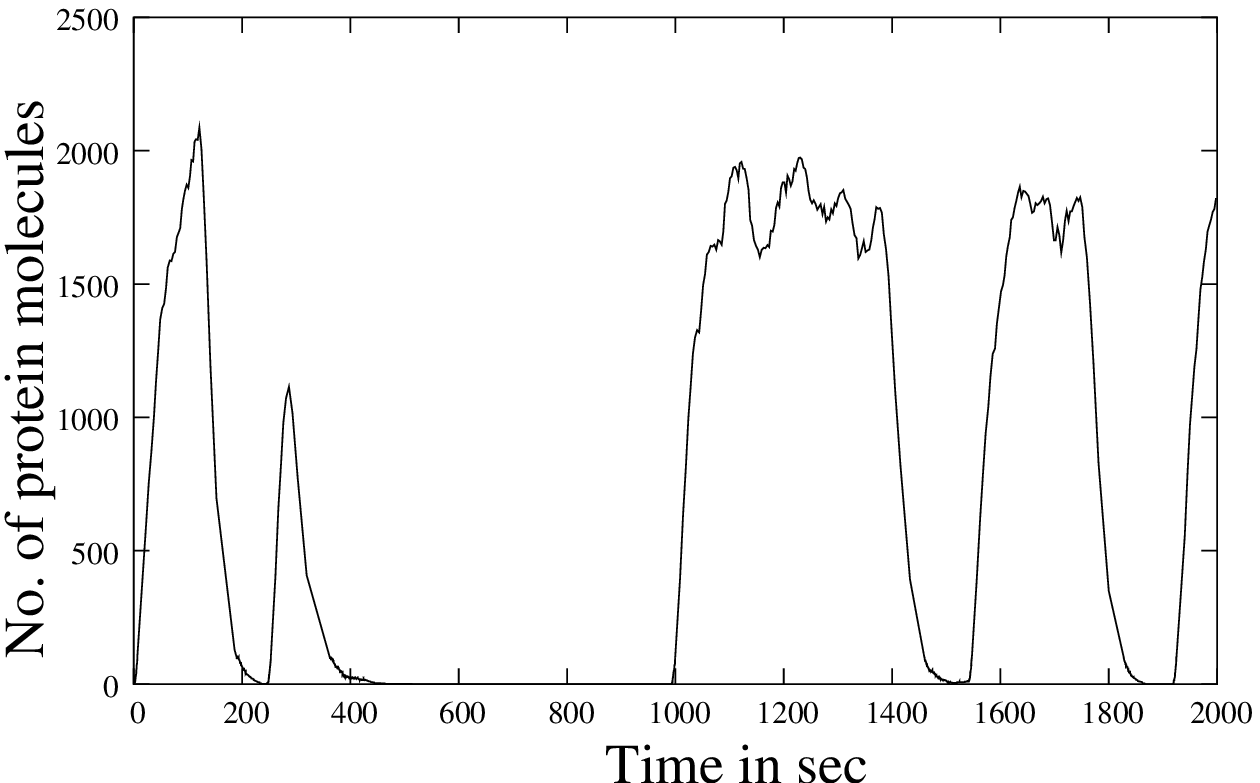}} \par}

FIG. 13(h). No. of protein molecules as a function of time. The stochastic
rate constants and the other parameter values are the same as in Fig.
7 except that \( c_{8}=0.015. \) 
\end{figure}

\begin{figure}[ihtp]
{\centering \resizebox*{2.85in}{!}{\includegraphics{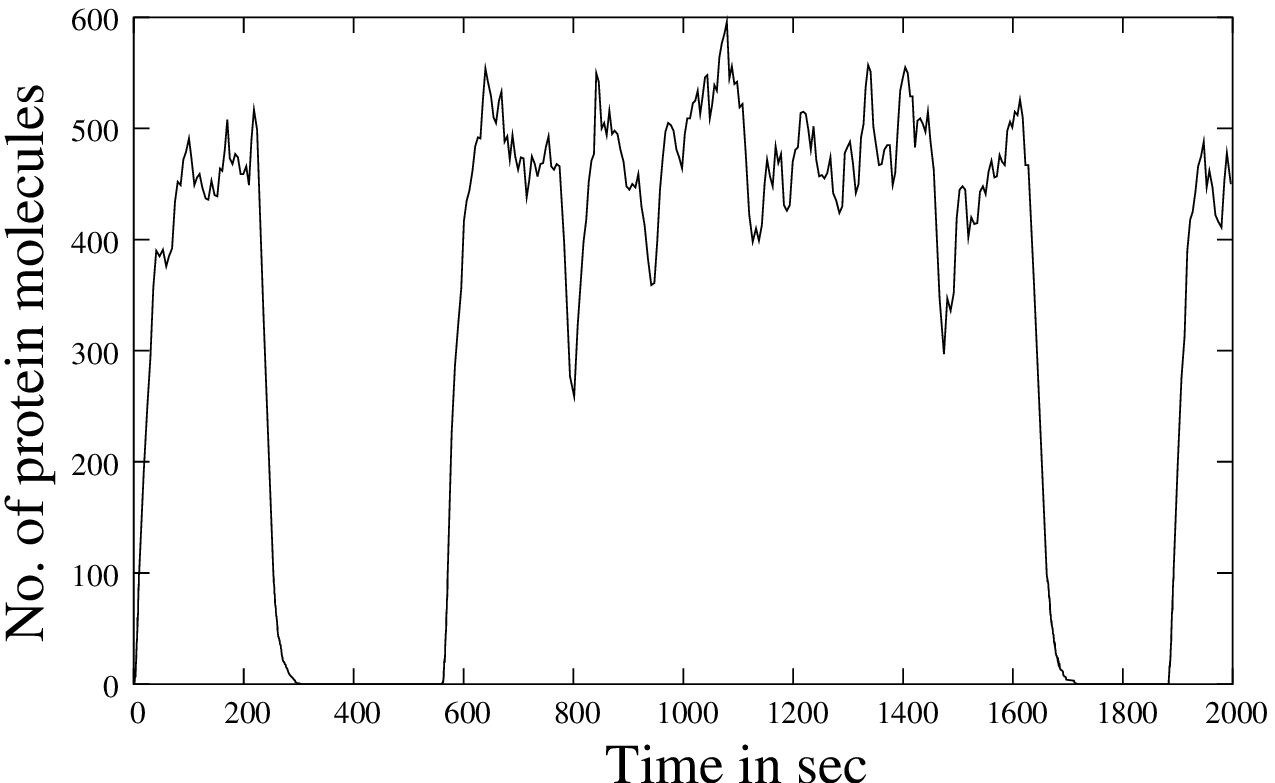}} \par}

FIG. 13(i). No. of protein molecules as a function of time. The stochastic
rate constants and the other parameter values are the same as in Fig.
7 except that \( c_{8}=0.004. \) 
\end{figure}

\begin{figure}[ihtp]
{\centering \resizebox*{2.85in}{!}{\includegraphics{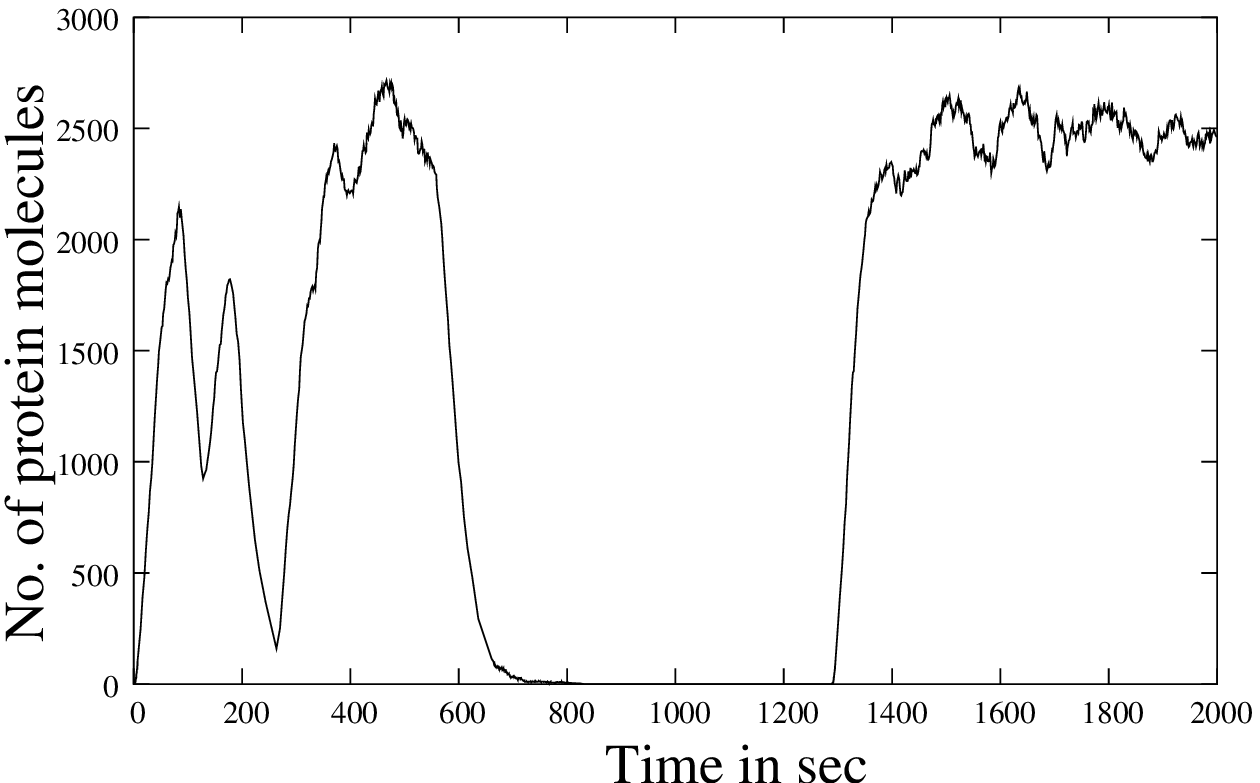}} \par}

FIG. 13(j). No. of protein molecules as a function of time. The stochastic
rate constants and the other parameter values are the same as in Fig.
7 except that \( c_{10}=0.1. \) 
\end{figure}
\onecolumn

We have also studied the effect of changing the number \( N_{R} \),
\( N_{RNAP} \) and \( N_{Rib} \) of R, RNAP and ribosome molecules
respectively. As \( N_{R} \) increases from 10 (Fig. 7), the Type
C pattern is gradually lost. Reducing \( N_{R} \) increases the total
duration of state 2. If\( N_{R} \)= 0, state 2 becomes the steady
state. If \( N_{RNAP} \) is reduced from 400 (Fig. 7) the Type C
pattern worsens gradually. If \( N_{RNAP} \) increases beyond 400,
the Type C pattern is still obtained. The magnitude of the protein
level in state 2 remains more or less the same. If \( N_{Rib} \)
decreases from 200, the total duration of state 2 increases.

Some general conclusions that can be made, on the basis of the stochastic
simulation of gene expression patterns, are as follows. The protein
level attained in state 2 is not affected by changes in the rate constants
c\( _{1} \), c\( _{2} \), c\( _{4} \), c\( _{7} \), c\( _{11} \),
c\( _{12} \) and the enhancement factor q. The rate constants c\( _{5} \),
c\( _{6} \), c\( _{8} \), c\( _{10} \) and c\( _{13} \) determine
the magnitude of the protein level. If c\( _{1} \) decreases and
c\( _{2} \) increases, the total duration T of state 2 in the time
interval 0-2000s increases. Transition from Type C \( \longrightarrow  \)Type
B can occur by decreasing c\( _{1} \)(Fig. 13(b)), increasing q and
by making the values of c\( _{5} \) and c\( _{6} \) unequal (Fig.
13(g)). Transition from Type C \( \longrightarrow  \) Type A pattern
can occur by increasing c\( _{1} \) (Fig. 13(d)), c\( _{8} \) (Fig.
13(h)) or by decreasing c\( _{10} \) (Fig.13(j)). Type C patterns
are favourable for high values of c\( _{5} \) and c\( _{6} \) with
c\( _{5} \)\( \simeq  \) c\( _{6} \) and also c\( _{11} \) \( \simeq  \)
c\( _{12} \). 

Kepler and Elston\cite{key-10} have proposed a simple mathematical
model of gene expression with no feedback. In their model, if an activator
molecule occupies the operator region, protein production occurs at
a rate \( \alpha _{1}. \) If the operator is unoccupied by the activator
molecule, protein production occurs at a lower rate \( \alpha _{0}. \)
The model does not include the intermediate steps of protein synthesis
like transcription, ribosome binding for the initiation of translation
etc. The model, however, provides lots of physical insight on stochastic
effects in the form of fluctuations in the discrete states of the
operator (unoccupied/occupied) on gene expression. Chemical reactions
that change the state of the operator are termed operator fluctuations.Approximations
to the dynamics were made for the cases in which the protein number
is large or the operator fluctuations are fast. In the first case,
the effective rate of protein synthesis fluctuates randomly in time
between high (synthesis rate \( \alpha _{1} \)) and low (synthesis
rate \( \alpha _{0} \)) levels. In the latter case, the fluctuations
are effectively averaged out over larger time scales. In our model,
the major biochemical reactions/events in protein synthesis have been
included and instead of activators we have regulatory molecules which
act as repressors. The simulation based on GA provides an accurate
knowledge of the microscopic origins of the different types of temporal
gene expression patterns. A simple mathematical model provides insight
on the origin of different patterns and transition from one type pattern
to another. Let \( m \) denote the number of proteins at time t devided
by the maximum number of proteins. The equation \begin{equation}
\label{mathed:tweentysecond-eqn}
\frac{dm}{dt}=x-m
\end{equation}
describes the rate of change in the number of proteins. The possible
values of \( x \) are \( 0 \) and \( 1 \) so that in the steady
state \( m \) can be either \( 1 \) (high level, state 2) or \( 0 \)
(low level, state 1). The variable \( x \) randomly switches between
the two states and the transition rate from state 1 \( \rightarrow  \)
state 2 is \( r_{1} \) and that from state 2 \( \rightarrow  \)
state 1 is \( r_{2} \). Let \( P_{j}(m=M,t) \), \( j=1,2 \), be
the probabilities of being in the state \( j \). One can then write
down the master equations\cite{key-34}

\begin{equation}
\label{mathed:tweentythird-eqn}
\frac{\partial P_{0}}{\partial t}=-\frac{\partial (-MP_{0})}{\partial t}-r_{1}P_{0}+r_{2}P_{1}
\end{equation}

\begin{equation}
\label{mathed:tweentyfourth-eqn}
\frac{\partial P_{1}}{\partial t}=-\frac{\partial ((1-M)P_{1})}{\partial t}-r_{2}P_{1}+r_{1}P_{0}
\end{equation}

The steady state distribution of \( P(m,t)=P_{0}(m,t)+P_{1}(m,t) \)
is given by 

\begin{equation}
\label{mathed:tweentyfifth-eqn}
P(m=M)=AM^{r_{1}-1}(1-M)^{r_{2}-1}
\end{equation}
where \( A \) is the normalization constant. Figs. 14(a), (b) and
(c) show the distribution \( P(m) \) versus \( m \) for different
values of \( r_{1} \) and \( r_{2} \). The distribution in Fig.
14(a) corresponds to the Type C pattern of gene expression (the magnitude
of the high protein level in state 2 is normalized to the value \( 1 \))
in the steady state and is bimodal in nature. Fig. 14(b) describes
the Type B pattern of gene expression and there is a single peak corresponding
to the steady state level, the magnitude of which is normalized to
\( 1 \). Fig. 14(c) shows a broad distribution in protein levels.
Fig. 12 shows a pattern of gene expression based on simulation of
the detailed model which gives rise to a probability distribution
of protein levels similar to that shown in Fig. 14(c). The transition
rates \( r_{1} \) and \( r_{2} \) are functions of the different
stochastic rate constants though the actual functional relationship
is yet to be worked out. The rate \( r_{1} \) depends dominantly
on \( c_{2}, \) \( c_{5}, \) \( c_{6}, \) \( c_{8} \) and \( c_{10} \)
whereas the rate \( r_{2} \) is mainly determined by \( c_{1} \)
and \( c_{13}. \) The simple model illustrates how the different
probability distributions arise and the transition from one type to
another occurs as the transition rates \( r_{1} \) and \( r_{2} \)
are changed. For slow transition rates, the states 1 and 2 can be
distinguished and the probability distribution \( P(m) \) is bimodal.
For fast transition rates, \( m \) has values intermediate between
\( 0 \) and \( 1 \) and the bimodality is smeared into a single
broad distribution. For large \( r_{1} \) and small \( r_{2} \),
the peak is around the high value \( 1 \). The transitions from one
type of probability distribution to another as \( r_{1} \) and \( r_{2} \)
are varied are consistent with the changes in the nature of the temporal
patterns of gene expression (Figs.12 and 13), brought about by changes
in the various stochastic rate constants on which the transition rates
\( r_{1} \) and \( r_{2} \) of the simpler model depend.

\begin{figure}[ihtp]
{\centering \resizebox*{3.2in}{!}{\includegraphics{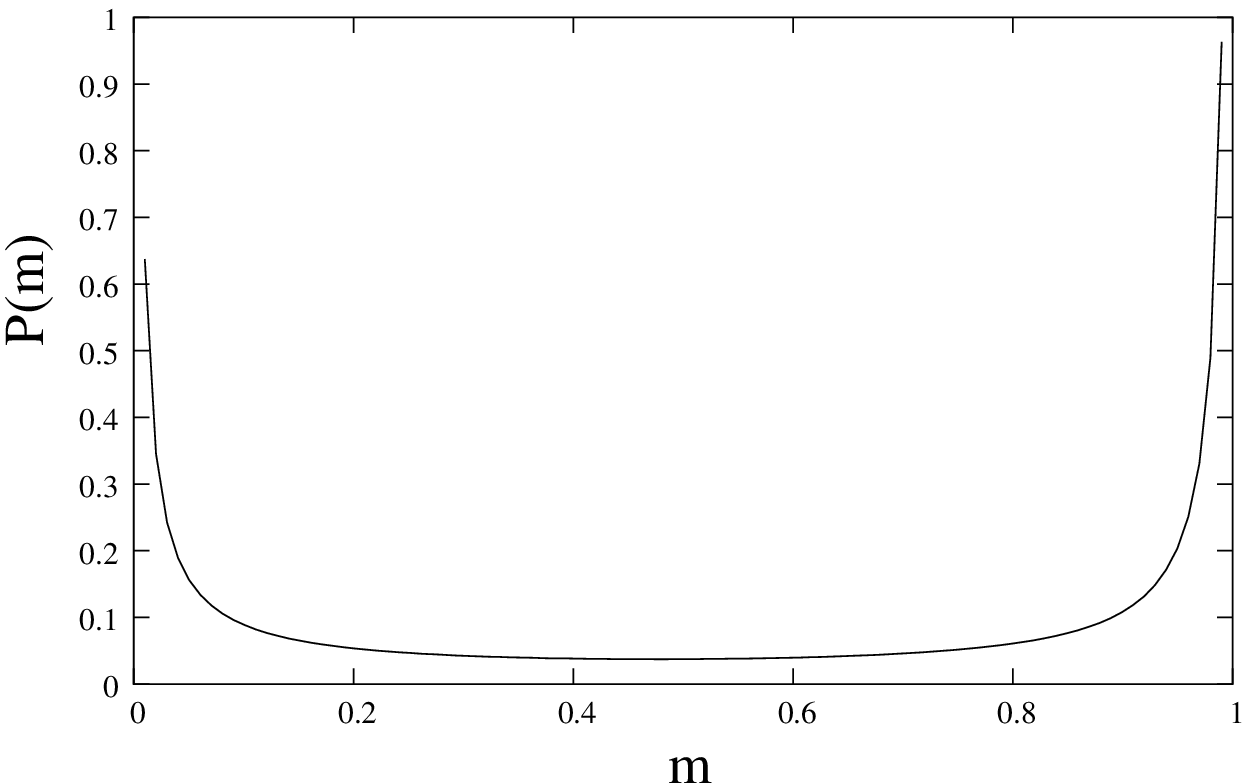}} \par}

FIG. 14(a). Distribution of \( P(m) \) as a function of \( m \)
for \( r_{1}=0.1, \) and \( r_{2}=0.01. \) 
\end{figure}

\begin{figure}[ihtp]
{\centering \resizebox*{3.2in}{!}{\includegraphics{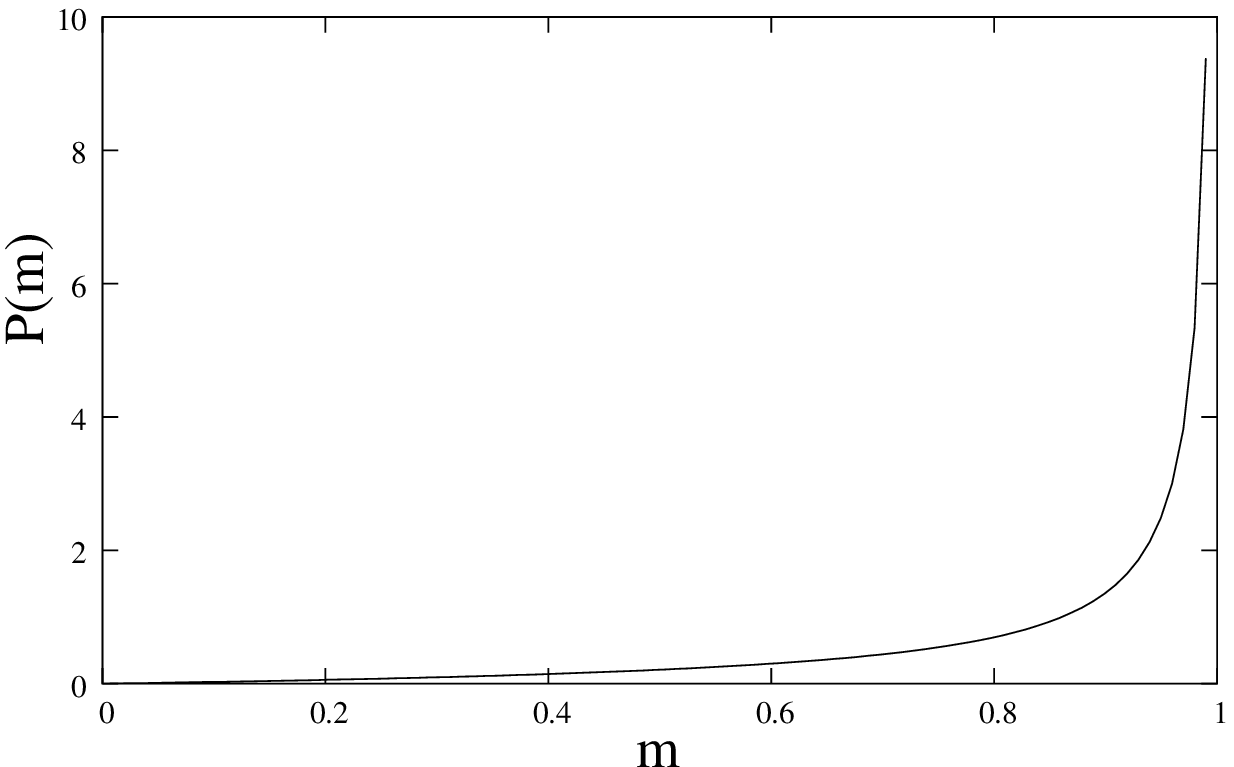}} \par}

FIG. 14(b). Distribution of \( P(m) \) as a function of \( m \)
for \( r_{1}=2.0, \) and \( r_{2}=0.2. \)
\end{figure}

\begin{figure}[ihtp]
{\centering \resizebox*{3.2in}{!}{\includegraphics{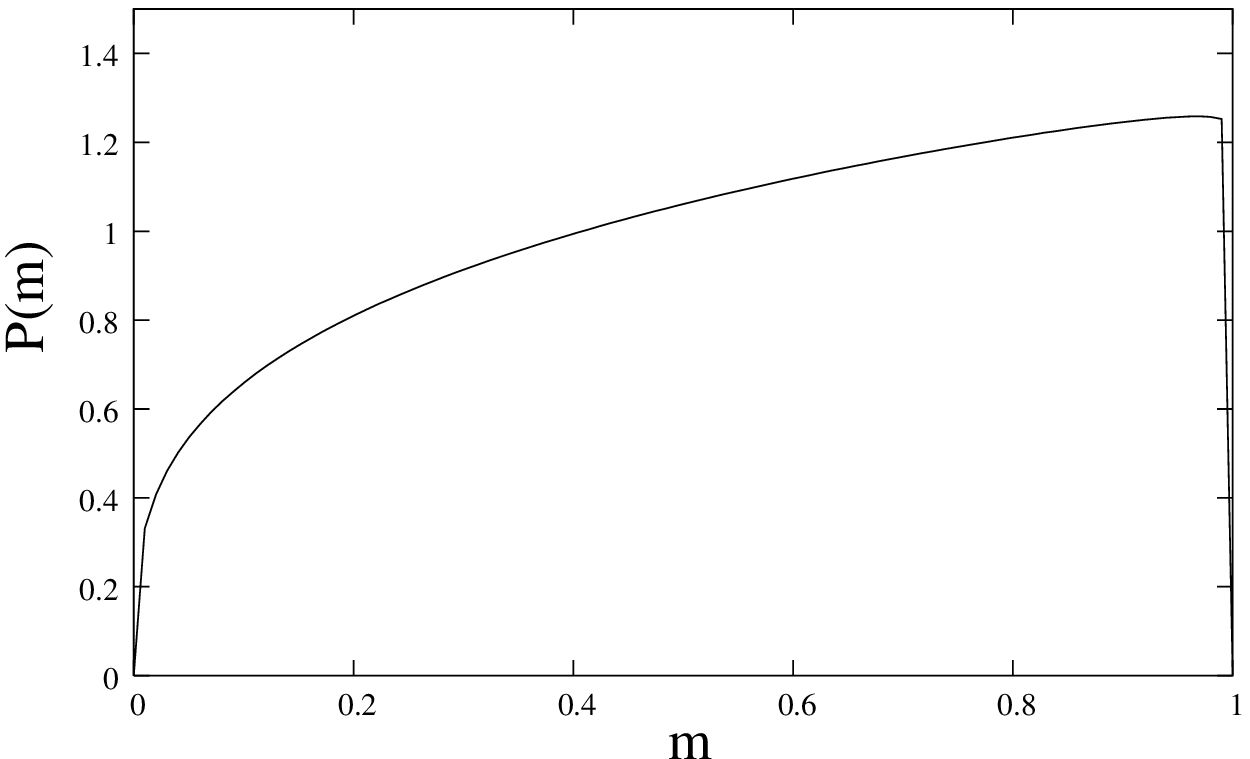}} \par}

FIG. 14(c). Distribution of \( P(m) \) as a function of \( m \)
for \( r_{1}=1.3, \) and \( r_{2}=1.01. \)
\end{figure}

\section*{IV. Concluding remarks}

In this paper, we have considered two models of gene expression in
a single cell. In the first model, the effect of inducer molecules
is explicitly considered and the cellular kinetics are described by
a set of sixteen biochemical reactions. In the second model, the cellular
kinetics are described by a set of thirteen biochemical reactions.
Both the models include the basic steps of transcription and translation
as well as transcriptional regulation through the binding of a regulatory
molecule R to the operator region, with R acting as a repressor of
transcription. Additionally, in the second model a transcriptionally
generated cooperativity factor q in the binding of RNAP has been introduced
as an optional feature. The patterns of gene expression as a function
of time have been determined in both the models with the highly accurate
stochastic simulation method based on the Gillespie Algorithm. Analytic
approaches \cite{key-10,key-11,key-12,key-13} are possible in studying
the temporal evolution but in these cases the detailed biochemical
reactions are lumped together into a few effective processes. This
makes it possible to determine the temporal evolution of the system
in a chemical Master Equation approach which is stochastic in nature.
Alternatively, the effect of stochasticity can be taken into account
by the inclusion of {}``noise'' terms in the differential rate equations.
The GA provides a more detailed picture of the kinetics though the
computational efforts required in its implementation are considerable.
In section II, we have studied gene expression in the presence of
inducer molecules. The major result of this study is that when the
number of inducer molecules, \( N_{I} \), is greater than or equal
to the number of regulatory molecules, \( N_{R} \), the cellular
steady state is state 2 in which the proteins are synthesized at a
high level. When \( N_{I}\ll N_{R} \) , the cellular state is state
1 in which the protein level is low/zero. As \( N_{I} \) approaches
\( N_{R} \), protein production occurs at intermediate levels. Fig.1
provides an example of how drastic the effect of even a single molecule
can be on gene expression. There is a considerable difference of protein
levels attained in the steady state when \( N_{I} \) is changed from
\( 19 \) to \( 20 \). This single molecule effect can be observed
over a wide region of parameter space and may give rise to threshold
phenomena. These results are new and have not been reported earlier.
McAdams and Arkin\cite{key-2} have pointed out that even a single
molecule can switch the biochemical state of a cell. Togashi and Kaneko
\cite{key-60} have studied an autocatalytic reaction system with
a small number of molecules and shown that due to the nonlinear dynamics,
amplification of small changes can give rise to single molecule switches.

The results of our simulation explain why an autocatalytic induction
mechanism gives rise to the {}``all-or-none'' phenomenon observed
in cells in the presence of inducer molecules\cite{key-22,key-23,key-24}.
Due to autocatalytic induction, the small number of inducer molecules
initially present in a cell is quickly amplified so that the condition
\( N_{I}\geq N_{R} \) is satisfied in a short interval of time. The
cell then exists in state 2 with a high protein level. Other cells
in which \( N_{I} \) is zero exist in state 1 with zero protein level.
In the presence of a subsaturating concentration of inducer molecules,
the distribution of protein levels in an ensemble of cells is bimodal.
In the absence of autocatalytic induction, protein production occurs
at low, high as well as at intermediate levels so that the bimodal
distribution gets smeared. This is in keeping with experimental observations. 

The model studied in section III does not include inducer molecules.
Simulation based on the GA shows the existence of three types of pattern
of gene expression as a function of time. In the Type A pattern, protein
synthesis occurs in abrupt stochastic bursts and a variable number
of protein is produced in each burst. There is considerable experimental
evidence \cite{key-16,key-18,key-19} for this type of gene expression
and in section III, some examples of this type of pattern have been
given (Figs.3 and 4). In Type B pattern, protein levels reach a steady
state (Figs. 5 and 6). This is very common type of pattern observed
in gene expression experiments.

In the Type C pattern (Fig. 7), the cellular state makes random transitions
between states 1 and 2 . This is a different manifestation of the
{}``all or none'' phenomenon and measurement levels in an ensemble
of cells is bimodal (Fig. 8). Type C patterns of gene expression have
been obtained by Kepler and Elston \cite{key-10} in their study of
model gene expression systems (see Fig. 4 of Ref. \cite{key-10})
using the Master Equation Approach. In these models detailed biochemical
reactions have been replaced by a few effective processes so that
the models are mathematically tractable. Kepler et al. have specifically
considered the effect of fluctuations in the state of the operator
on gene expression. Their conclusion is that the operator fluctuations
can induce bistability in parameter regions which gives rise to monostability
in the deterministic, i.e., the zero noise limit or destroy bistability
if it exists in the noise-free case. In the deterministically bistable
region, the gene acts like a genetic switch and external noise/perturbation
is needed to flip the switch from one state to the other. This is
the principle of operation behind the noise-based switches and amplifiers
for gene expression proposed by Hasty et al \cite{key-13}. If a system
is stochstically bistable, the fluctuations in the system flip the
switch between the two states (say, state 1 and state 2) at random
time intervals. As in Ref. \cite{key-10}, the operator fluctuations
have been explicitly considered in the models studied in Sections
II and III and it has been shown that in certain parameter regimes,
stochastically bistable behaviour corresponding to Type C pattern
of gene expression is obtained. Our stochastic simulation results,
obtained by using the highly accurate Gillespie Algorithm, provide
a microscopic basis as well as quantitative estimates of the different
rate constants for obtaining Type C pattern of gene expression. The
simulation method can be used to study a large number of reactions
which is not feasible in the formalism of a mathematical model. The
significant omission in Ref. \cite{key-10} is that no distinction
has been made between transcription and translation. The models describe
direct translation from gene into protein. As pointed out in the paper,
the simplification may have considerable impact on cellular phenomena.
For example, the synthetic repressilator network \cite{key-8} would
not oscillate if a time delay between transcription and translation
were absent. In our model of gene expression, all the major biochemical
reactions involved in transcription and translation have been explicitly
taken into account and the time delay is clearly seen in Fig. 11.
There is some experimental evidence \cite{key-31,key-57,key-59,key-32}
of transcriptionally generated cooperative binding of RNAP to the
promoter region. For non-zero valus of the cooperativity factor \( q \)
in our model the duration of state two (high protein level) is found
to increase. The result is new and the effect of cooperative RNAP
binding on gene expression needs to be investigated in greater detail.

We have further studied the effect of changing the various stochastic
rate constants on the temporal pattern of gene expression. Transitions
involving Type C \( \rightarrow  \) Type B and Type C \( \rightarrow  \)
Type A patterns of gene expression have been obtained by changing
appropriate stochastic rate constants. These transitions can be understood
in the framework of a simple mathematical model which does not include
the detailed biochemical reactions. The model describes a simple gene
expression system in which random transitions occur between states
1 and 2, corresponding to low and high levels of protein production.
The model has two parameters \( r_{1} \) and \( r_{2} \) which are
the transition rates from state 1 \( \rightarrow  \) state 2 and
state 2 \( \rightarrow  \) state 1 respectively. In the steady state,
the probability distribution of protein levels can be calculated.
A bimodal distribution corresponds to Type C patterns whereas an unimodal
distribution is obtained for Type B patterns. A broad distribution
in protein levels (Fig. 14(c)) is obtained corresponding to the gene
expression pattern shown in Fig. 12. Transition from one type of distribution
to another can be obtained by changing the rate constants \( r_{1} \)
and \( r_{2} \). While experimental evidence for Type A and Type
B patterns of gene expression is considerable, we do not know of specific
experiments exhibiting Type C patterns. As mentioned in the Introduction,
the Type C pattern is similar to that obtained in the case of two-state
jump processes in which transitions between two states occur at random
time intervals \cite{key-34,key-37}. An example is provided by a
spin-\( \frac{1}{2} \) in the presence of a magnetic field and in
contact with a heat bath \cite{key-37}. The literature on two state
jump processes is large and one measurable quantity of interest is
the mean first-passage time (MFPT). The time required to switch between
two states is a random variable and is known as the first first passage
time. Determination of the MFPT and other characteristic measures
of the two-state jump processes describing Type C gene expression
has not been attempted in this paper. The Type C pattern is reminiscent
of a binary digital pulse with states 1 and 2 corresponding to the
{}``0'' (OFF) and {}``1'' (ON) states. Genes with expression pattern
of Type C may be combined together to construct binary logical circuits.
Recently, synthetic networks of genes displaying features of binary
logic circuits have been constructed \cite{key-38}. Bialek \cite{key-12}
has studied stability and noise in biochemical switches and has shown
that switches with long periods of stability and switchability in
milliseconds can be constructed from fewer than a hundred molecules.
The conclusion is arrived at by studying a model of the synthesis
of a single biochemical species, say, proteins in the Langevin formalism.
The result obtained is of considerable interest but needs to be verified
in a detailed approach involving intermediate processes. The correlation
of the amount of random variation in protein distribution, measured
by the Fano factor, with the transcriptional and translational rates
can be determined in the stochastic simulation approach and the results
compared with those obtained in experiments \cite{key-20,key-21}.
A detailed study on the dominant contributions to noise in our models
of gene expression is in progress and the results will be reported
elsewhere. 

{\centering \textbf{ACKNOWLEDGEMENTS}\par}

I. B. acknowledges helpful discussions with A. M. Kierzek. R. K. was
supported by the Council of Scientific and Industrial Research, India
under Sanction No. 9/15 (239) / 2002 - EMR - 1.

\end{document}